\documentclass[11pt,a4paper]{article}
\pdfoutput=1
\usepackage{amsfonts}
\usepackage{jheppub}
\usepackage{amsmath}
\usepackage{amssymb}
\usepackage{color}
\usepackage{graphicx}%
\usepackage{float}
\usepackage[normalem]{ulem}
\usepackage{mathrsfs}
\usepackage[scr=rsfs]{mathalfa}
\newcommand{\stkout}[1]{\ifmmode\text{\sout{\ensuremath{#1}}}\else\sout{#1}\fi}
\allowdisplaybreaks

\title{\boldmath Asymptotic analysis of energy functionals in anti-de Sitter spacetimes}

\author[a]{Giorgos Anastasiou,}
\author[b,c]{Mart\'in Bravo,}
\author[d]{Rodrigo Olea}

\affiliation[a]{Departamento de Ciencias, Facultad de Artes Liberales, Universidad Adolfo Ibáñez, \\
Avenida Diagonal Las Torres 2640, 7941169, Peñalolén, Santiago}
\affiliation[b]{Instituto de Ciencias Exactas y Naturales, Universidad Arturo Prat, \\
Playa Brava 3256, 1111346, Iquique, Chile}
\affiliation[c]{Facultad de Ciencias, Universidad Arturo Prat, \\
Avenida Arturo Prat Chac\'on 2120, 1110939, Iquique, Chile}
\affiliation[d]{Instituto de Física, Pontificia Universidad Católica de Valparaíso,\\
Casilla 4059, Valparaíso, Chile.}

\emailAdd{georgios.anastasiou@uai.cl, martbravo@estudiantesunap.cl, rodrigo\_olea\_a@yahoo.co.uk}

\abstract{
Conformal Gravity (CG) is a Weyl--invariant metric theory whose action is free from divergences for generic asymptotically anti-de Sitter spaces.  For Neumann boundary conditions, it reduces to renormalized Einstein--AdS gravity at tree level. By evaluating CG's action on a replica orbifold, one obtains a codimension-2 local conformal invariant functional, $L_\Sigma$, which reduces to the renormalized area, the reduced Hawking mass and the Willmore Energy, for a surface $\Sigma$. Although there is evidence supporting the idea that this functional should be finite, a detailed analysis of its asymptotic behavior near the conformal boundary is still lacking. In this work, the finiteness of the conformal surface functional $L_\Sigma$ is shown for any boundary--anchored surface embedded in an arbitrary ambient spacetime which is a solution to CG.  This conclusion is drawn regardless  the fact the surface is minimal or not. This result implies that Conformal Renormalization method not only applies to the bulk action, but also codimension--2 functionals.
}
\begin{document}
\maketitle
\flushbottom
\section{Introduction}
\label{sec:intro}

Energy functionals describe the mechanical deformations of surfaces immersed in $\mathcal{M}^3$ and have relevant applications in physics and biology.
For example, the Willmore energy, used as a model for red blood cells and lipid bilayers, like vesicles, is defined for a closed smooth $2$--dimensional surface $\Sigma$ with genus $g$ embedded in $\mathbb{R}^3$. 
In the context of gravitational physics, energy functionals play a central role in capturing key geometric and physical properties of embedded surfaces.

For example, in 1973, Penrose~\cite{penrose1973naked} claimed that in any asymptotically flat spacetime which contains a black hole, the total mass is always bounded below by $\sqrt{\mathcal{A} / 16 \pi}$, where $\mathcal{A}$ is the area of its event horizon. 
This result assumes a series of global properties of Einstein spaces,  which  are closely related to the Cosmic Censorship conjecture.  Cosmic Censorship states that the formation of spacetime singularities is possible only if there is an horizon that protects them.
Therefore, the proof of the Penrose inequality (PI) is an open problem of utmost relevance within the framework of gravitational collapse.  
In that respect, it was later shown by Huisken and Ilmannen~\cite{huisken2001inverse} that PI holds for the case of a single black hole with a $3$--dimensional time--symmetric initial data that obeys the dominant energy  condition. 
In order to prove this, they use the inverse mean curvature flow (IMCF), a geometric evolution equation where the velocity of a spacelike surface is inversely proportional to its mean curvature~\cite{geroch1973energy}. 
An important ingredient in this derivation is  the Hawking mass, which is a quasilocal definition of mass for a $2$--dimensional closed surface $\Sigma$~\cite{Hawking:1968qt}.
Then the inequality follows from the fact that this functional is monotonically increasing under a generalized form of IMCF  and approaches asymptotically to the Arnowitt-Deser-Misner (ADM) mass. 
Shortly afterwards, Bray generalized the previous result for the multiple black holes case using the so--called conformal flow~\cite{bray2001proof}. 
This was later extended up to eight dimensions in Ref.\cite{Bray:2007opu}. 
(See Ref.~\cite{Mars:2009cj} for an interesting overview on the subject). 

From a holographic standpoint, plenty of applications of energy functionals can be found in the context of the anti-de Sitter (AdS) {gravity}/ Conformal Field Theory (CFT) correspondence~\cite{Maldacena:2014blg, gubser1998gauge, witten1998anti}. This duality between a supergravity theory in asymptotically anti-de Sitter (AAdS) spacetimes and a CFT in one dimension lower becomes manifest by 
the matching of the corresponding partition functions which establishes a map connecting observables on the field theory side to dynamical degrees of freedom on gravity.
As a consequence, the holographic dictionary provides a geometrical interpretation of the different field theoretic observables. 

Surface functionals also appear in the computation of entanglement entropy (EE) of a CFT. 
EE is a measure of the quantum correlation between two or more subsystems and it is a key quantity to study phase transitions and critical phenomena~\cite{Calabrese:2004eu, Klebanov:2007ws}, characterization of topological phases of matter~\cite{Kitaev:2005dm, Levin:2006zz, Papanikolaou:2007}, quantum information~\cite{Horodecki2009, Nielsen_Chuang_2010} and black hole thermodynamics~\cite{Harlow:2014yka, Solodukhin:2011gn, Emparan:2006ni}.
{In Refs.~\cite{ryu2006holographic, ryu2006aspects}}, Ryu and Takayanagi (RT) proposed that the EE of a spatial region $A$ is given by the area of a minimal codimension--$2$ surface $\Sigma$ embedded in the bulk and homologous to $A$, i.e.,
\begin{align}
    S(A) = \frac{\mathcal{A}\left( \Sigma \right)}{4 G_N} \, . \label{RTformula}
\end{align}
The RT conjecture was later proven by Lewkowycz and Maldacena (LM) in Ref.~\cite{lewkowycz2013generalized}. 
This was achieved by extending the replica trick to the gravity action, generalizing the gravitational entropy to configurations with discrete symmetry. 
The discrete symmetry induces identifications which give rise to a conical defect in the bulk manifold. When evaluating the gravity action in the resulting orbifold, one ends up with a codimension--$2$ functional which resides at the singularity.

In Einstein-AdS gravity, LM prescription identifies the conical contribution of the corresponding action as the area of the surface $\Sigma$, i.e., the RT formula.
However, as $\Sigma$ is anchored to the conformal boundary, its area inherits the infrared divergencies of the AdS volume. 

Therefore, a proper renormalization prescription for the area functional is an absolute must. In Ref.~\cite{taylor2016renormalized}, the authors, using standard holographic renormalization techniques, determined a series of intrinsic counterterms for the area of $\Sigma$. An extension of this procedure can be found in Ref.~\cite{Taylor:2020uwf}. An alternative description, based on the addition of topological terms to the bulk action, was provided in Refs.~\cite{ anastasiou2018topological,Anastasiou:2018mfk}. This proposal makes explicit the connection to the notion of Renormalized Area introduced by Alexakis and Mazzeo in Ref.~\cite{alexakis2010renormalized}. Renormalized Area correctly isolates the finite part of holographic EE for minimal surfaces. However, the latter statement is no longer true for non-minimal surfaces due to divergencies coming from the non-orthogonal intersection to the boundary.

A generalization of the Hawking mass in the presence of asymptotically hyperbolic spaces was introduced in Ref.~\cite{fischetti2017bound} by Fischetti and Wiseman, dubbed reduced Hawking mass. Taking advantage of its monotonicity under an IMCF flow for AAdS manifolds the authors showed that the finite part of the EE of an arbitrary region in $\left(2+1\right)$--dimensional holographic CFTs admits an upper bound depending on the energy density of the state.

It should be stressed at this point that corrections to Renormalized Area are unavoidably ad hoc in Einstein gravity. Indeed, a well-posed variational principle should imply that classical configurations are stationary under arbitrary variations of the spacetime and surface functionals.  Thus, while the Einstein equations hold in the bulk, in codimension--$2$ the minimality condition is necessarily met.

In that respect, additional extrinsic curvature terms in the surface functional indicate the departure of the classical conditions on $\Sigma$. While quadratic-curvature terms may be considered as an alternative starting point for the bulk, the requirement of matching codimension--$2$ functionals for Einstein gravity imposes several constraints in the form of that action. This construction is the subject of the next section.

\section{Quadratic Curvature Gravity and conical defects} \label{section 2}

Energy functionals can be derived using the LM method, which is a systematic prescription to obtain codimension--$2$ quantities of holographic interest, from the conical contribution of gravitational bulk actions.

\subsection{Lewkowycz--Maldacena Prescription} \label{subsec2.1}

In Quantum Field Theory (QFT), the EE of a spatial region $\mathcal{A}$ of a $d$--dimensional manifold $C$  is commonly calculated in the path integral formulation using the replica trick~\cite{Calabrese:2004eu}. 
This method involves the construction of an $n$--fold cover $C_n$ by joining the boundaries of $n$ replicas of the region boundary, with a cut along the entangling surface.
As a consequence, the manifold $C_n$ acquires a $\mathbb{Z}_n$ symmetry, corresponding to the cyclic permutation of the replicas when rotating by an angle $2\pi (1-1/n)$.
This symmetry induces an orbifold $C^{(n)} = C_n/ \mathbb{Z}_n$, which is characterized by a conical singularity along the entangling surface. 
After analytically continuing the replica parameter $n$, the EE is given by
\begin{equation}
    S(\mathcal{A}) =  - \lim\limits_{n \rightarrow 1} {n \partial_{n} \left( \ln{Z_n} - n \ln{Z_1} \right)} \, ,
\end{equation}
where $Z_n$ is the partition function of $C^{(n)}$, and $Z_1 = Z$ is the partition function of the original manifold $C$.

While the replica trick has been extensively applied to study entanglement entropy in diverse QFT cases -- such as in non-interacting QFTs -- it is often very difficult to obtain exact results~\cite{casini2009entanglement,Kitaev:2005dm}.
For strongly coupled CFTs, the AdS/CFT correspondence provides a geometric interpretation of EE on the gravity side, which allows for computations directly from the gravitational action. 
The computational power of gauge/gravity duality is unfolded by relating the partition function of the CFT to the on--shell Euclidean action $I_{\text{grav}}$ of the gravitational dual~\cite{gubser1998gauge, witten1998anti}
\begin{equation}
\ln{Z_{\text{CFT}}} \simeq - I_{\text{grav}}  \, . 
\end{equation}
Ref.~\cite{lewkowycz2013generalized} established a direct connection between EE and the gravitational action, by evaluating the latter on a manifold $\mathcal{M}$ with a squashed conical singularity.
In this approach, the $n$--fold cover $C_n$ on the boundary is dual to a regular Euclidean $n$--fold cover $\mathcal{M}_n$ in the bulk. 
Assuming the $ \mathbb{Z}_n$ symmetry holds for the bulk geometry $\mathcal{M}_n$, one can construct an orbifold $\mathcal{M}^{(n)} = \mathcal{M}_n /  \mathbb{Z}_n$.
The bulk on--shell action for this orbifold is then related to that of the original manifold by the relation  $I[\mathcal{M}_n]  = n {I}\left[ \mathcal{M}^{(n)} \right]$.
Thus, in the holographic context, the entanglement entropy of a region $A$ can be computed as
\begin{equation}
    S(A) = - \left. {\partial_{\vartheta}  I \left[ \mathcal{M}^{\left( \vartheta \right)} \right]} \right|_{\vartheta = 1} \, , \label{lwformula}
\end{equation}
where $\vartheta = 1/n$ is the inverse replica parameter, and the orbifold geometry has an angular deficit of $2 \pi (1 - \vartheta)$. 
On general grounds, the bulk action evaluated on an orbifold can be decomposed into a regular and a singular contribution.  In the case of Einstein gravity, the corresponding action takes the form
\begin{equation} \label{Einstein in an orbifold}
    \frac{1}{16 \pi G_N}\int\limits_{\mathcal{M}^{\left(\vartheta\right)}}d^4 X \sqrt{g} R^{\left(\vartheta\right)} = \frac{1}{16 \pi G_N} \int\limits_{\mathcal{M}}d^4 X \sqrt{g} R + \frac{(1 - \vartheta)}{4 G_N}\mathcal{A}\left[ \Sigma \right] \, ,
\end{equation}
where 
\begin{equation}
    \mathcal{A}\left( \Sigma \right) = \int\limits_{\Sigma} d^2 Y \sqrt{\gamma}\, ,
\end{equation}
is the area of the codimension--$2$ surface $\Sigma$ located at the conical singularity, where the induced metric is denoted as $\gamma_{ab}$, with the indices $(a,b, \cdots)$ referring to the intrinsic coordinates. 
Indeed, the first term corresponds to the Einstein--Hilbert action, evaluated on a smooth manifold $\mathcal{M}$, while the second term can be identified as the Nambu--Goto action which described a codimension--$2$ cosmic brane with tension $T = \frac{(1 - \vartheta)}{4G}$, embedded in the spacetime geometry ~\cite{Dong:2013qoa}. 
By substituting this expression into Eq.\eqref{lwformula} and taking the tensionless limit $\vartheta = 1$, the well--known Ryu--Takayanagi formula \eqref{RTformula} for EE is recovered for a minimal surface $\Sigma_{min}$. 

The LM prescription can be extended to gravitational theories, which include higher-curvature corrections~\cite{Dong:2013qoa, Camps:2013zua, miao2014}. 

In the next section, we examine a modification to Einstein gravity in the form of quadratic terms in the curvature and explore the codimension--$2$ functional derived from this class of theories. Particularly remarkable is the case where the quadratic couplings in the curvature produce Conformal Gravity in the bulk, which is the only theory which is invariant under Weyl rescaling of the metric. For that particular theory, the LM procedure unveils the connection between a Weyl invariant in the bulk with another residing on a $2$--dimensional surface. As was shown in Ref.\cite{anastasiou2022energy}, this codimension--$2$ Weyl invariant can be linked to energy functionals of interest in AdS gravity with applications within the gauge/gravity framework.

\subsection{Quadratic Curvature Gravity}
Quadratic curvature gravity is one of the simplest modifications to the Einstein--Hilbert action and naturally arises within the context of effective field theory~\cite{buchbinder1992effective, Donoghue:1995cz, Salvio:2018crh}.
Its action contains quadratic terms in the curvature tensor, which serve to regularize the ultraviolet divergences from a given quantization scheme applied to Einstein Gravity~\cite{Stelle:1977}.

In an arbitrary dimension, a proper basis of curvature-squared terms can be given by $Rie^2$, $Ric^2$ and $R^2$ contributions.
In turn, in four dimensions, as the Gauss--Bonnet (GB) term $\mathcal{E}_4 \equiv \sqrt{g} \left( Rie^2 - 4 Ric^2 + R^2 \right)$ is a topological invariant, the most general action for QCG can be written as
\begin{equation}
    I_{QCG} = \frac{1}{16\pi G_N}\int\limits_{\mathcal{M}} d^4 X \sqrt{g} \left(\alpha R^{2} + \beta {{Ric}}^{2} \right) \, , \label{QCG}
\end{equation}
where $\alpha$ and $\beta$ are coupling constants. 
The variation of the action with respect to the metric yields,
\begin{align}
    \mathcal{E}^{\mu \nu} \equiv {}& \beta \Box \left( R^{\mu \nu} - \frac{1}{2}R g^{\mu \nu} \right) + 2 \beta \left( R^{\mu \sigma \nu \rho} - \frac{1}{4}g^{\mu \nu} R^{\sigma \rho} \right)R_{\sigma \rho}  +  2 \alpha R\left( R^{\mu \nu} - \frac{1}{4} g^{\mu \nu} R \right) \nonumber \\ 
{}& + \left( 2 \beta + \alpha \right) \left( g^{\mu \nu} \Box - \nabla^{\mu} \nabla^{\nu} \right) R = 0 \label{EOM-QCG} \, ,
\end{align}
which is a set of fourth--derivative differential equations. 
In general, QCG can suffer from Ostrogradsky instabilities~\cite{Ostrogradsky:1850fid, Woodard:2015zca} due to its high--derivative nature, which leads to ghost--like degrees of freedom.
As a matter of fact, the only combination of curvature-squared terms that still leads to second-order field equations is the Gauss-Bonnet term (in higher dimensions).
\newline
Einstein-AdS spaces, defined by 
\begin{equation}\label{Einstein spaces}
    R_{\mu \nu} = -  \frac{3}{\ell^2} g_{\mu \nu} \, ,
\end{equation}
where $\ell$ is the AdS radius, can be alternatively expressed in terms of a vanishing traceless Ricci tensor 
\begin{equation}\label{Traceless Ricci}
    H_{\mu \nu} \equiv R_{\mu \nu} - \frac{1}{4} R\,  g_{\mu \nu}\, .
\end{equation}
These spaces are solutions not only to General Relativity with negative cosmological constant but also to Eq. \eqref{EOM-QCG}.
Furthermore, when the action \eqref{QCG} is evaluated on this class of solutions, it becomes proportional to the volume of the 
corresponding AAdS spacetime
\begin{align}
    I_{\rm{QCG}} \left[ \text{E} \right] = \frac{36}{\ell^2} \left( 4 \alpha + \beta \right) {\rm{Vol}} \left( \mathcal{M}\right) \, . \label{VolQCG}
\end{align}
This quantity exhibits a divergent behavior, which is the reflection of a conformal structure at the boundary.\\
In the context of gauge/gravity duality, a standard method to remove these divergences in Einstein-AdS gravity is holographic renormalization, which introduces a series of counterterms constructed from the boundary metric~\cite{Henningson:1998gx, deHaro:2000vlm, Bianchi:2001kw, Skenderis:2002wp}. 

On general grounds, for AAdS solutions of QCG, the asymptotic resolution of the field equations order by order in the holographic coordinate would also dictate the expression for the boundary counterterms and the form of the corresponding renormalized action $I^{ren}_{QCG}$.  However, a renormalization scheme which requires the addition of  boundary terms in a given spacetime foliation does not lend itself to the direct use of the LM method. Indeed, working out the contribution of conical singularities on the boundary terms in $I^{ren}_{QCG}$ would be quite involved, as it amounts to the projection of the counterterms on the codimension--$2$ surface.\footnote{This procedure was implemented for Einstein gravity in Ref.\cite{taylor2016renormalized}. In practice, it amounts to the calculation of the first counterterms of the series in order to renormalize the area functional.} 

An alternative approach, which considers the addition of extrinsic counterterms at the boundary was proposed for Einstein-AdS gravity in Refs.~\cite{Olea:2005gb,Olea:2006vd}. The form of the boundary structures in this prescription remains the same regardless of the inclusion of higher-curvature terms in the action, as in Lovelock gravity~\cite{Kofinas:2008ub} and Einstein-QCG theories~\cite{PhysRevD.101.064046}. Only the overall factor in front and the effective AdS length change accordingly. 
In four spacetime dimensions, this renormalization method amounts to the addition of the second Chern form $B_3$ 
\begin{equation}
    I^{ren}_{\text{QCG}} =  \frac{1}{16\pi G_N} \left[ \int\limits_{\mathcal{M}} d^4 X \sqrt{g} \left(\alpha R^{2} + \beta {{Ric}}^{2} \right) +\gamma \int\limits_{\partial \mathcal{M}} d^{3} x B_3  \right] \, 
    \, , \label{IQCG_1}
\end{equation}
   which, in a compact notation, can be written as
\begin{equation}
B_3 =  4 \sqrt{h}\,\delta^{ j_1 j_2 j_3}_{i_1 i_2 i_3} \mathscr{K}^{i_1}_{j_1} \left( \frac{1}{2} \mathscr{R}^{i_2 i_3}_{j_2 j_3} - \frac{1}{3} \mathscr{K}^{i_2}_{j_2} \mathscr{K}^{i_3}_{j_3} \right) \, ,
\end{equation}
where we consider the radial foliation of $\mathcal{M}$.
In this context, the metric in Gauss normal coordinates is given by
\begin{equation}
 ds^2 = N^{2}(z) dz^2 + h_{ij}(z,x) dx^i dx^j \, ,
\end{equation}
where $\{ x^i \}$ are the coordinates of the boundary $\partial \mathcal{M}$ and $h_{ij}$ is the induced metric at a fixed value of $z$. The induced metric defines the intrinsic Riemann tensor $\mathscr{R}^{i}_{jkl}$ and the extrinsic curvature is expressed as
\begin{equation}
 \mathscr{K}_{ij} = - \frac{1}{2N} \partial_{z} h_{ij} \, .
\end{equation}
The second Chern form enters the Gauss--Bonnet theorem for non--compact manifolds,  
\begin{equation}
    \int\limits_{\mathcal{M}} d^4 X \mathcal{E}_4 = 32 \pi^2 \chi(\mathcal{M}) + \int\limits_{\partial \mathcal{M}} d^3x B_3 \, ,
\end{equation}
as it may be thought as the boundary correction to the Euler characteristic $\chi(\mathcal{M})$. 
The above formula shows that the boundary contribution is locally equivalent to the Gauss--Bonnet density, such that the renormalized action~\eqref{IQCG_1} can be expressed in terms of bulk quantities. As a requirement,  $\mathcal{E}_4$ should at least cancel the divergences in the volume  for global AdS spacetime. This argument fixes the GB coupling such that the total action is
\begin{align}
    I^{ren}_{QCG} = {}& \frac{1}{16 \pi G_N} \left[ \int\limits_{\mathcal{M}} d^4 X \sqrt{g} \left(\alpha R^{2} + \beta {{Ric}}^{2} \right) - \frac{3}{2}\left( 4 \alpha + \beta \right)  \int\limits_{\mathcal{M}} d^4 X \mathcal{E}_4 \right] \nonumber  \\
    {}&  + \frac{3 \pi}{G_N}\left( 4 \alpha + \beta \right) \chi(\mathcal{M}) \, . \label{IQCG_2}
\end{align}
In what follows, the corresponding codimension-2 functional from this renormalized version of QCG action is explicitly worked out.

\subsection{Renormalized Area in QCG}
The contribution coming from conical defects in the evaluation of  quadratic-curvature actions were systematically studied by Fursaev, Patrushev and Solodukhin (FPS) in Ref.~\cite{fursaev2013distributional}. Smoothing out the apex of the cone by a distribution function, upon a proper limit, makes explicit the geometric functional in two dimensions lower in terms of the intrinsic and extrinsic curvatures of the surface $\Sigma$. Equivalently, the FPS relations are expressed in terms of projections of the bulk Riemann tensor, i.e.,
\begin{itemize}
    \item Riemann-squared term:
\begin{align}\label{Riem conical}
    & \int\limits_{\mathcal{M}^{\left( \vartheta \right)}} d^4 X \sqrt{{g}} \left(Rie^{\left( \vartheta \right)}\right)^{2} = 
    \int\limits_{\mathcal{M}} d^4 X \sqrt{{g}}\,  Rie ^{2} \nonumber \\ 
    & + 8 \pi \left( 1 - \vartheta \right) \int\limits_{\Sigma} d^2 Y \sqrt{{\gamma}} \left( R_{\lambda \mu \sigma \rho }n_{A}^{\lambda }n^{\sigma A}n_{B}^{\mu }n^{\rho B} - \mathcal{K}^{A}_{ ab } \mathcal{K}_{A}^{ab} \right) + O \left((1- \vartheta)^2 \right) \, ,
\end{align}
    \item Ricci-squared term:
\begin{align}\label{Ricci conical}
      & \int\limits_{\mathcal{M}^{\left( \vartheta \right)}} d^4 X \sqrt{{g}} \left(Ric^{\left( \vartheta \right)}\right)^{2} =
    \int\limits_{\mathcal{M}} d^4 X \sqrt{{g}} Ric^{2} \nonumber \\
    & + 4 \pi \left( 1- \vartheta \right) \int\limits_{\Sigma} d^2 Y \sqrt{{\gamma}} \left( R_{\mu \rho }n_{B}^{\mu }n^{\rho B} - \frac{1}{2} \mathcal{K}_{A} \mathcal{K}^{A} \right) +  O \left( \left( 1- \vartheta \right)^2 \right) \, ,
\end{align}
    \item Ricci scalar-squared term:
\begin{equation} \label{Rscalar conical}
     \int\limits_{\mathcal{M}^{\left( \vartheta \right)}} d^4 X \sqrt{{g}} {\left(R^{\left( \vartheta \right)}\right)}^{2} = 
    \int\limits_{\mathcal{M}} d^4 X \sqrt{{g}} R^{2} + 8 \pi \left( 1-\vartheta \right) \int\limits_{\Sigma} d^2 Y \sqrt{{\gamma}} \left( R \right) +  O \left( \left(1- \vartheta \right)^2 \right) \, .
\end{equation}
   \end{itemize}
Here, $\mathcal{R}$ is the Ricci scalar associated to the surface metric $\gamma_{ab}$ and $n^{\mu}_{A}$ are the normal vectors to the surface. Capital letters $(A, B, \cdots)$ denote the orthogonal directions and $n^{\mu}_{A}n^{B}_{\mu}=\delta^{B}_{A}$ is the metric of the normal bundle.
The extrinsic curvature $\mathcal{K}^{A}_{ab}$ expressed the embedding of $\Sigma$ in $\mathcal{M}$.

In the previous subsection, the chosen basis of quadratic terms in the curvature involves the GB term. Therefore, it is convenient to see the implication of the above formulas for this topological invariant, which leads to
\begin{equation}
    \int\limits_{\mathcal{M}^{\left( \vartheta \right)}} d^4 X \,\mathcal{E}_4^{\left( \vartheta \right)} = \int\limits_{\mathcal{M}} d^4 X \mathcal{E}_4 + 8\pi \left( 1-\vartheta \right) \int\limits_{\Sigma} d^2 Y \,\mathcal{E}_2 \, .
\end{equation}
Notice that, in the last formula, there are no higher-order corrections in the conical deficit, as a consequence of topological protection of invariants of the Euler class.

Then, the renormalized QCG action \eqref{IQCG_2}, evaluated on an orbifold $\mathcal{M}^{(\vartheta)}$ decomposes into a regular and a singular part, i.e.,
\begin{align}
    {\left(I_{\text{QCG}}^{\text{ren}}\right)}^{\left( \vartheta \right)} = I_{QCG}^{ren} + \frac{(1 - \vartheta)}{4 G_N} L_{QCG} \left[ \Sigma \right] + \mathcal{O}\left( \left( 1 - \vartheta \right)^2 \right) \, ,
\end{align}
where $L_{QCG} \left[ \Sigma \right]$ is the conical contribution at first order, whose explicit form is given by
\begin{align}\label{generalL}
  L_{QCG} \left[ \Sigma \right] = {}& \int\limits_{\Sigma}d^2 Y \sqrt{\gamma} \left[ - 3 \left( 4 \alpha + \beta \right) \mathcal{R}  + \beta  R_{\mu \nu }n_{A}^{\mu }n^{\nu A} + 2 \alpha  R - \frac{\beta}{2} \mathcal{K}^{A} \mathcal{K}_{A}  \right]  \nonumber \\ 
   {}&  + 12 \pi \left(  4 \alpha + \beta  \right) \chi \left(\Sigma \right)  \, .
\end{align}
The appearance of the  Euler characteristic of the surface is a direct consequence of the relation
\begin{equation}
      \chi\left(\mathcal{M}^{\left( \vartheta \right)}\right) =  \chi \left( \mathcal{M} \right) + \left( 1-\vartheta \right) \chi \left( \Sigma \right) \, .
\end{equation}
As discussed above, minimal surfaces play an essential role in holographic applications. In Einstein gravity, they are defined by the extremization of the area functional under arbitrary variations along any of the normal vectors $n^A$ to the surface. In that case, the minimality condition implies the vanishing of the trace of the extrinsic  curvature
\begin{equation} \label{minimal condition}
     \mathcal{K}^{A} = 0\,. 
\end{equation}
As an abuse of language, in what follows, the last condition will be a defining feature of minimal surfaces in QCG, although extremization of the codimension-2 functional \eqref{generalL} would lead to a more general equation.

Einstein spacetimes \eqref{Einstein spaces} are a consistent sector of QCG. For an Einstein ambient space and a minimal surface $\Sigma_{\rm{min}}$, the generic functional \eqref{generalL} reduces to
\begin{equation}
     L_{QCG} \left[ \Sigma_{\text{min}}, \text{E} \right] = - \frac{6 \left( 4 \alpha + \beta \right)}{\ell^2} \mathcal{A}_{\rm{ren}} \, ,
\end{equation}
what is a generalization  of the notion of renormalized area for a  minimal surface in Einstein--AdS gravity   ~\cite{alexakis2010renormalized}
\begin{equation}
    \mathcal{A}_{\rm{ren}} = \frac{\ell^2}{2} \int\limits_{\Sigma_{min}} d^2 Y \sqrt{\gamma} \left( \mathcal{R} + \frac{2}{\ell^{2}} \right) - 2 \pi \ell^2 \chi \left(\Sigma \right) \, . \label{renorareatopological}
\end{equation}
In Einstein gravity, this surface functional can be obtained by applying the LM method directly to the Einstein--Hilbert action, renormalized by adding a topological term~\cite{Anastasiou:2017xjr}. It readily removes divergences from the conformal boundary for surfaces anchored in it orthogonally (minimal surfaces). 

\section{Codimension-2 functionals from QCG}\label{section 3}
In this section, a couple of properties of the codimension-2 functional~\eqref{generalL} are used as a selection criterion of the corresponding gravity theory in the bulk from a general QCG.

\subsection{Non-minimal surfaces and codimension-2 functionals}

In Einstein gravity, the finiteness of the Renormalized Area \eqref{renorareatopological} is spoiled when one departs from minimality condition     \eqref{minimal condition}. Indeed, the anchoring points of non-minimal surfaces, which intersect the conformal boundary at an arbitrary angle, generate new divergent contributions from the IR sector of AAdS gravity.

In Ref.~\cite{fischetti2017bound}, Fischetti and Wiseman proposed a different energy functional, the so-called reduced Hawking mass for Einstein--AdS spaces, which includes a correction due to the extrinsic curvature, that is,

\begin{equation} \label{RHM}
    I_H \left( \Sigma \right) \equiv 2 \int\limits_{\Sigma}d^2 Y\sqrt{\gamma}\left( \mathcal{R} + \frac{2}{\ell^2} - \frac{1}{2} \mathcal{K}^{A}\mathcal{K}_{A} \right) \, .
\end{equation}
This functional is monotonic under an inverse mean curvature flow and it is finite for generic surfaces.

Having in mind the relations \eqref{Riem conical}--\eqref{Rscalar conical}, it is quite clear that a term of the type $\mathcal{K}^A\mathcal{K}_A$ in Eq.\eqref{RHM} cannot result from the application of the LM method to neither Einstein gravity nor topological terms in the action. In contrast, the quadratic terms in the curvature in QCG produce the generic surface functional
\begin{equation} \label{L in Einstein}
    L_{QCG} \left[ \Sigma,\text{E} \right] = - 3 \left( 4 \alpha + \beta \right) \left[ \int\limits_{\Sigma} d^2 Y \sqrt{\gamma} \left( \mathcal{R} + \frac{2}{\ell^2} + \frac{\beta}{6 \left( 4 \alpha + \beta  \right)} \mathcal{K}^{A} \mathcal{K}_{A}\right) - 4 \pi  \chi \left(\Sigma \right) \right] \, ,
\end{equation}
when evaluated on Einstein ambient spaces. 
To obtain a conical contribution from QCG related to the reduced Hawking mass, one must impose the relation $\alpha = - \beta/3$ in Eq.~\eqref{L in Einstein}. Under this condition, $L_{QCG}$ reduces in the Einstein sector to the form
\begin{equation} 
    L_{QCG} \left[ \Sigma,\text{E} \right] = \beta \left( \frac{1}{2}I_{H} - 4 \pi \chi(\Sigma) \right)\, ,
\end{equation}
with an undetermined overall factor.
At the level of the bulk action, this specific combination of coupling constants leads to
\begin{align}
    I_{CG}  = {}& \beta \left[  \frac{1}{32 \pi G_N} \int\limits_{\mathcal{M}} d^4 X \sqrt{{g}} \, \left( Rie^2 - 2 Ric^2 + \frac{1}{3}R^2 \right) -  \frac{ \pi }{G_N} \chi(\mathcal{M}) \right]
    \, , \label{QCG as CG}
\end{align}
which can be identified with the action of Conformal Gravity (CG) in four dimensions. With the interest of rendering Weyl invariance manifest, it is convenient to express this gravity action in terms of the Weyl tensor
\begin{equation}
     W^{\alpha \beta}_{\mu \nu} = R^{\alpha \beta}_{\mu \nu} - \left( S^{\alpha}_{\mu} \delta^{\beta}_{\nu} - S^{\beta}_{\mu} \delta^{\alpha}_{\nu} - S^{\alpha}_{\nu} \delta^{\beta}_{\mu} + S^{\beta}_{\nu} \delta^{\alpha}_{\mu} \right) \, , 
\end{equation}
where $S_{\mu \nu}$ is the Schouten tensor
\begin{equation}
S_{\mu \nu} = \frac{1}{2} \left( R_{\mu \nu} - \frac{1}{6}R \, g_{\mu \nu} \right) \, ,
\end{equation}
which plays the role of a compensator field of the Riemann tensor under Weyl rescalings of the metric. In doing so, the Lagrangian is cast in the following form
\begin{equation}
     W_{\mu \nu}^{\alpha \beta}W^{\mu \nu}_{\alpha \beta} = Rie^2 - 2 Ric^2 + \frac{1}{3}R^2\,.
\end{equation}
In the next subsection, the same combination of quadratic-curvature couplings in the bulk leads to a symmetry enhancement in the codimension-2 functional from diffeomorphic to Weyl invariance. This is understood by the fact that conical defects in Conformal Gravity induce yet another Weyl invariant in two dimensions lower, for an arbitrary spacetime geometry.

\subsection{Weyl Invariance on conical defects}

Weyl invariants are geometric objects that remain unchanged under Weyl rescalings of the metric. The number of independent Weyl invariants increases with dimension, and their explicit form is known only up to 8 dimensions~\cite{Bonora:1985cq,Deser:1993yx,Boulanger:2007ab,Erdmenger:1997gy,Boulanger:2004zf,Boulanger_2022}. In the bulk, they may be thought of as a theory of modified, higher-derivative gravity. At the boundary of odd-dimensional AdS gravity, conformal invariants appear in the type B conformal anomaly, whose coefficients (central charges)  provide a criterion for identifying the corresponding boundary CFT. In the same spirit, one may construct Weyl invariant objects defined in submanifolds, which depend on  both intrinsic and extrinsic quantities. In particular, in two-dimensional submanifolds, there are two independent objects that transforms homogeneously under local Weyl rescalings, which can be cast in the form
\begin{equation}
  P^{A}_{ab}P^{ab}_{A} \, , \qquad  W_{ab}^{ab} \, .
\label{weylinvariantscod2}
\end{equation}
Here, $P^{A}_{ab}$ is the traceless part of the extrinsic curvature, defined by
\begin{equation}
  P^{A}_{ab} = \mathcal{K}^{A}_{ab} - \frac{1}{2} \mathcal{K}^{A} \gamma_{ab} \, ,
\end{equation}
and $W_{ab}^{ab}$ is the pullback of the Weyl tensor, which can be written as
\begin{equation}\label{Weylproyection}
W_{ab}^{ab} =  R_{\lambda \mu \sigma \rho }n_{A}^{\lambda }n^{\sigma A}n_{B}^{\mu }n^{B\rho } -R_{\mu \rho }n_{B}^{\mu }n^{\rho B} +\frac{1}{3}R  \, .
\end{equation}
Notice that this basis consists on second-derivative quantities. As a consequence, only four-derivative bulk objects, in the form of quadratic-curvature couplings, may be considered as a starting point for LM method. The natural question is whether, for given values of $\alpha$ and $\beta$ in the QCG action~\eqref{IQCG_2}, the resulting codimension-2 functional is a Weyl invariant.

In particular, when  the scalar Gauss relation \eqref{GaussScalarRelation} is considered, then the action of Eq.~\eqref{generalL} can be decomposed as
\begin{equation} 
    L_{QCG} \left[ \Sigma \right] =  -3\left(4 \alpha + \beta \right)  L_{QCG}^{C}  - 2  \left(  3  \alpha + \beta \right)  L_{QCG}^{NC}\,.
\end{equation}
Here $L_{QCG}^{C}$ and $L_{QCG}^{NC}$ are the Weyl and non--Weyl invariant pieces, respectively, given by the following expressions 
\begin{equation}
    L_{QCG}^{C}\left[ \Sigma \right] = \int\limits_{\Sigma} d^2 Y \sqrt{\gamma}    \left(  W^{ab}_{ab} - P^{A}_{ab} P^{ab}_{A} \right)  - 4 \pi \chi \left(\Sigma \right)
    \label{LSigma}
\end{equation}
and
\begin{align} 
    L_{QCG}^{NC}\left[ \Sigma \right] =  \int\limits_{\Sigma} d^2 Y \sqrt{\gamma}    \left(  R - 2  R_{\mu \rho }n_{A}^{\mu }n^{\rho A} + \mathcal{K}^{A} \mathcal{K}_{A} \right)\, .
\end{align}
Firstly, notice that the latter vanishes identically when $  \alpha = - \beta / 3$, that is the same condition that leads to the reduced Hawking mass for Einstein spacetimes. Secondly, the LM prescription uniquely fixes the relative factor of the Weyl invariant combination to the one appearing in the GW anomaly~\cite{Graham:1999pm}. 

As a consequence, constructing 2D functionals with manifest Weyl invariance using the LM prescription uniquely fixes both the bulk action, i.e. conformal gravity, and the exact form of the functional. This property makes CG particularly relevant since it provides the only framework in which Weyl symmetry governs not only the bulk action but also extends to codimension--$2$ functionals.

\section{Conformal Gravity}\label{section 4}
The reasoning in the previous section singles out CG from the class of quadratic-curvature theories. Indeed, the criterion of an energy functional as proportional to the reduced Hawking mass necessarily leads to CG as a sensible theory in the bulk. In a similar fashion, Weyl invariance of the 2D functional is an additional, compelling argument which supports the same claim. In what follows, some interesting features of CG are rendered manifest.

\subsection{Finiteness of the bulk action}
The CG action is given by the expression
\begin{equation}
    I_{CG} = \beta \left[  \frac{1}{32 \pi G_N} \int\limits_{\mathcal{M}} d^4 X \sqrt{{g}} \, W_{\mu \nu}^{\alpha \beta}W^{\mu \nu}_{\alpha \beta}  -  \frac{ \pi }{G_N} \chi(\mathcal{M}) \right]
    \, . \label{CG action}
\end{equation}
As pointed out above, the presence of the Euler characteristic is required by the matching with different energy functionals in the literature.
A remarkable property of CG is that it is free of  infrared divergences from the conformal boundary of AAdS spacetimes without  additional counterterms~\cite{grumiller2014conformal}. This can be seen by usingn a power-counting argument in the holographic (radial) coordinate. In this regard, it is convenient to write down the metric for an AAdS space in the Fefferman--Graham (FG) gauge~\cite{fefferman1985conformal},
\begin{equation} 
    ds^2 = \frac{\ell^2}{z^2} dz^2 + \frac{1}{z^2} \bar{g}_{ij}(z, x) dx^{i} dx^{j} \, , \label{FG gauge}
\end{equation}
where the conformal boundary is located at $z=0$, and $ \bar{g}_{ij}(z, x)$ admits an expansion in terms of powers of the radial holographic coordinate
\begin{equation}
    \bar{g}_{ij}(z, x) = {g}^{(0)}_{ij}(x) + {z} {g}^{(1)}_{ij}(x) + {z^{2}} {g}^{(2)}_{ij}(x) + z^3g_{ij}^{(3)}(x) + \mathcal{O}\left( z^4 \right) \, . \label{eq.fefferman graham expansion}
\end{equation}
The mode $g^{(0)}_{ij}$ represents metric at the conformal boundary  and is the source for the holographic stress tensor.
The subleading term $g^{(1)}_{ij}$ is a consequence of bulk field equations which are of higher derivative in the radial coordinate. Therefore, it does not appear in Einstein--AdS gravity. 
Indeed, in CG this coefficient is a new holographic source, associated to a partially massless response in the holographic stress tensor~\cite{Grumiller:2013mxa}. 
The presence of this mode implies deviations from Einstein--AdS gravity and allows for additional freedom in the asymptotic structure of the metric a linear term in $z$ may eventually modify the asymptotic behavior of the spacetime curvature. 
In the FG gauge, the CG Lagrangian can be decomposed as
\begin{equation}
    W^{\alpha \beta}_{\mu \nu} W^{\mu \nu}_{\alpha \beta} = W^{ij}_{mn} W_{ij}^{mn} + 4 W^{i z}_{m n} W_{i z}^{m n} + 4 W^{i z}_{j z} W_{i z}^{j z} \, ,
\end{equation}
where $W^{ij}_{mn}$, $W^{iz}_{mn}$, and $W^{iz}_{jz}$ are the independent projections of the Weyl tensor. Their asymptotic expansion is 
\begin{align}
W^{iz}_{jz} = {}& \frac{z^2}{2\ell^2} \left[
- {H}^{(0)i}_{j} - \left( g^{(2)i}_{j} - \frac{1}{3}g^{(2)} \delta^{i}_{j} \right)
+ \frac{1}{4}  \left( g^{(1)i}_{j} - \frac{1}{3} g^{(1)} \delta^{i}_{j} \right) \, {g}^{(1)}
\right] + {O}(z^3) \, ,  \\
W^{iz}_{mn} = {}& \frac{z^2}{2\ell^2} \Bigg[
2 {D}^{(0)}_{[n} {g}^{(1)i}_{m]} 
+ \delta^i_n {D}^{(0)}_{[m} {g}^{(1)k}_{k]} - \delta^i_m {D}^{(0)}_{[n} {g}^{(1)k}_{k]}
\Bigg] + \mathcal{O}(z^3)\, ,  \\
W^{ij}_{mn} = {}& \frac{z^2}{2\ell^2} \Bigg[
\frac{1}{4} {g}^{(1)}_{kl} {g}^{(1)kl} \delta^{ij}_{mn}
- \frac{1}{12} \left({{g}^{(1)}}\right)^{2} \delta^{ij}_{mn}
+ {g}^{(1)} {g}^{(1)[i}_{[
m} \delta^{j]}_{n]} 
- 2\delta^{[i}_{[m} {g}^{(1)j]}_{k} {g}^{(1)k}_{n]} 
- \frac{1}{4} {g}^{(1)[i}_{[m} {g}^{j]}_{(1)n]} \nonumber \\
{}& - \frac{2}{3} {g}^{(2)} \delta^{ij}_{mn} + 4 \delta^{[i}_{[m} {g}^{(2)j]}_{n]} - 4  \delta^{[i}_{[m}\mathscr{R}^{(0)j]}_{n]} + \frac{1}{3} \mathscr{R}^{(0)} \delta^{ij}_{mn} + 2 \mathscr{R}^{(0)ij}_{mn} \Bigg] 
+ \mathcal{O}(z^3) \, .
\end{align}
Here, $D_{(0)i}$ is the covariant derivative associated to ${g}_{ij}^{(0)}$. Indices are raised and lowered with the same metric.\\
In the Einstein sector of CG, the last component of the bulk Weyl tensor at $\mathcal{O}(z^2)$ is proportional to the Weyl tensor of the metric at the conformal boundary, which vanishes identically.
As it can be seen from the above relations, for a generic AAdS space, each term of the CG Lagrangian falls off as $z^4$ or faster. 
This asymptotic behavior cancels the divergences from the determinant of the metric, $\sqrt{g} \sim z^{-4} \sqrt{{g}^{(0)}}$. As a consequence, the action remains finite,
\begin{equation}
    I_{CG} \sim \int_{\epsilon}^{z} dz' \int\limits_{\partial \mathcal{M}} d^3 x \frac{\sqrt{{g}^{(0)}}}{z'^{4}}z'^4 = \text{finite} + \mathcal{O}(\epsilon) \, ,
\end{equation}
where $\epsilon$ is the cutoff distance where the conformal boundary is located. This power-counting argument shows that CG is free of IR divergences for AAdS spacetimes. The full holographic analysis of CG can be seen in Ref.~\cite{Grumiller:2013mxa}.

Furthermore, the above result suggests that Einstein sector of CG should inherit this finiteness property. The latter reasoning goes along the line of Maldacena's argument concerning the equivalence between four-dimensional CG  and renormalized Einstein-AdS gravity at tree level when Neumann boundary conditions are imposed ~\cite{Maldacena:2011mk}. Alternatively, the cancellation of higher-derivative modes may be derived from taking Einstein  condition on the bulk Ricci tensor Refs.~\cite{Anastasiou:2016jix,Anastasiou:2020mik}. In particular, the connection between CG and the renormalized Einstein-AdS gravity becomes manifest, when the Weyl tensor is decomposed as
\begin{equation}
W_{\mu\nu}^{\alpha\beta}=W^{\alpha \beta}_{\mu \nu}[E]-X_{\mu\nu}^{\alpha\beta} \,,
\label{Weyltensordecomp}%
\end{equation}
where
\begin{align}
    W^{\alpha \beta}_{\mu \nu}[E] = R^{\alpha \beta}_{\mu \nu} + \frac{1}{\ell^2} \delta^{\alpha \beta}_{\mu \nu} 
    \, , \label{Weyl in Einstein}
\end{align}
and the $X$ tensor is given by
\begin{equation}
X_{\mu\nu}^{\alpha\beta}=\frac{4}{D-2}H_{[\mu}^{[\alpha}\delta_{\nu]}^{\beta
]}+\frac{1}{D\left(  D-1\right)  }\left[  \frac{D\left(  D-1\right)}{\ell^{2}}+R\right]
\delta_{\mu\nu}^{\alpha\beta}\,. \label{Xcontrib}%
\end{equation}
Here
\begin{equation}
H_{\mu\nu}=R_{\mu\nu}-\frac{1}{D}R g_{\mu\nu} \,,
\label{TracelessRicci}%
\end{equation}
is the traceless part of the Ricci tensor, that vanishes identically for Einstein spacetimes.
The fact that the CG field equations are of higher order in the radial derivative implies that the next-to-lead and the holographic coefficients in FG expansion~\eqref{eq.fefferman graham expansion} remain undetermined. The consistency with the Einstein limit of the theory requires the vanishing of $H^{\mu}_{\nu}$. Asymptotically, this is achieved by imposing: i) $ \partial_{z} \left. \bar{g}_{ij} \right|_{z = 0} = 0 $, ii) $\text{Tr}\left( \partial_{z}^{3} \bar{g}_{ij}\right) = 0$ and the identification of the mode $g^{(2)}_{ij}$ with the Schouten tensor of the boundary metric $g^{(0)}$ ~\cite{Anastasiou:2020mik}.

Since the Weyl tensor coincides with the AdS curvature tensor $W^{\alpha \beta}_{\mu \nu}[E]$ only when the Einstein condition is satisfied, the CG action \eqref{CG action} with the coupling constant $\beta =  {\ell^2}/{2}$ is equivalently rewritten as
\begin{equation}
    I_{CG}[E] = \frac{1}{16 \pi G} \int\limits_{\mathcal{M}} d^4 X \sqrt{{g}} \left( R + \frac{6}{\ell^2} \right) + \frac{\ell^2}{64 \pi G} \int\limits_{\mathcal{M}} d^4 X \, \mathcal{E}_4  -  \frac{ \pi \ell^2 }{2G_N} \chi(\mathcal{M})  \, . \label{topol action}
\end{equation}
This corresponds to the Einstein--AdS action renormalized by the addition of a topological term \cite{mivskovic2009topological}. As a consequence, embedding Einstein gravity in a Weyl-invariant theory to provide the counterterms necessary to cancel the divergences of AAdS spaces.
This property defines the Conformal Renormalization framework, which links finiteness of the action to a Weyl-invariant completion.
Evidence which underpins that claim has been given in the case of scalar--tensor theories~\cite{PhysRevD.107.104049,Barrientos:2022yoz} and in six--dimensional AdS gravity~\cite{Anastasiou:2020mik, Anastasiou:2023oro}.

\subsection{Penrose-Brown-Henneaux analysis in Conformal Gravity}\label{PBH}

Any AAdS spacetime admits a Taylor-like expansion in the radial coordinate $z$ for the metric $\bar{g}_{ij}$ 
of the form~\eqref{FG gauge}, which defines the FG frame \cite{fefferman1985conformal}. In a large class of higher-curvature gravity theories, including CG, the mode  $g_{\left(1\right)ij}$ can be consistently switched on. However, this is not the case in CG.
In the case of CG, all terms of the FG expansion up to the normalizable order remain dynamically undetermined. Nevertheless, there are kinematic arguments which allow to partially fix some of them. In particular, earlier work indicates the universal character of $g_{\left(2\right) ij}$ for any non-degenerate theory based on residual symmetries that leave invariant the asymptotic structure~\cite{Imbimbo:1999bj}. These are the Penrose--Brown--Henneaux (PBH) transformations, i.e., a subset of diffeomorphisms that preserves the FG gauge of the metric \cite{Brown:1986nw, Imbimbo:1999bj, Penrose:1985bww}
\begin{equation}
   z' = z + {\xi}^{z} (z,x) \, , \qquad x'^{i} = x^{i} +  {\xi}^{i} (z,x) \,.
\end{equation}
In what follows, the kinematic analysis given by the PBH transformations restricts the AdS asymptotics in higher-curvature gravity, i.e., for a non-vanishing $g_{\left(1\right) ij}$ mode. Keeping the FG form of the metric amounts to the conditions
\begin{equation}
\mathcal{L}_{\xi} g_{zz} = \mathcal{L}_{\xi} {g}_{iz} = 0 \, ,
\end{equation}
in terms of the Lie derivative along the vector $\xi^{\mu}$.
The above conditions are satisfied by diffeomorphisms that behave as 
\begin{align}
    \xi^{z}(z, x) =  {}& -  {z} \Omega({x})  \, , \label{xizeta}\\
    \xi^{i} (z, x)= {}&   \ell^{2}  \partial_{j} \Omega(x)  \int^{z}_{0} z' \bar{g}^{ij}(z',x) dz'  \, , \label{xi^i} \, 
\end{align}
where $\Omega(x)$ is an arbitrary function. Due to the asymptotic behavior of $\bar{g}_{ij}(z, x)$, the vector  $\xi^{i}(z,x)$ can, in turn, be expanded as
\begin{equation} \label{expansion of xi^i}
    \xi^{i}(z, x) = \sum^{\infty}_{n=2} z^{n} \xi^{(n)i} (x)  \, .
\end{equation}
As shown in Ref.~\cite{Imbimbo:1999bj}, for diffeomorphisms of the form of Eqs.~(\ref{xizeta}) and (\ref{xi^i}), the variation of $\bar{g}_{ij}(z, x)$ is given by
\begin{equation}
   \delta \bar{g}_{ij}(z, x) = \Omega \left( x \right) \left( 2 - z \partial_{z} \right) \bar{g}_{ij}(z, x) + \bar{D}_{i} \xi_{j}(z, x) + \bar{D}_{j} \xi_{i}(z, x)  \, ,
   \label{PBH variation of g}
\end{equation}
where $\bar{D}_{i}$ is the covariant derivative associated to $\bar{g}_{ij}$. After inserting Eqs.~\eqref{eq.fefferman graham expansion}, \eqref{expansion of xi^i}  and \eqref{PBH variation of g} in the last expression,  the transformation of each term of the FG expansion $g^{(n)}_{ij}$ is obtained, what reads
\begin{equation} \label{delta g}
    \delta g_{ij}^{(n)} = 
        \Omega \left( x \right) \left( 2 - n \right) g_{ij}^{(n)} + \sum_{r = 2}^{n} \left( g^{(n-r)}_{il} \partial_{j}\xi^{(r)l} + g^{(n-r)}_{jl} \partial_{i}\xi^{(r)l} + \xi^{(r)l} \partial_{l} g^{(n-r)}_{ij} \right) \, .
\end{equation}
As expected, the first term represents the fact that a radial diffeomorphism induces a Weyl rescaling at the conformal boundary since
\begin{equation}
 \delta {g}_{ij}^{(0)} =  2 \Omega (x) {g}_{ij}^{(0)} \,.
\end{equation}
It is interesting to notice that the same behavior is met by $g^{(1)}_{ij}$, with a different scaling factor
\begin{equation}
    \delta {g}_{ij}^{(1)} =  \Omega (x) {g}_{ij}^{(1)} \, .
\end{equation}
This result extends the PBH analysis of Ref.\cite{Imbimbo:1999bj} to an AAdS metric with higher-derivative modes. In an AdS/CFT framework, the fact PBH transformations produce a  Weyl rescaling in $g^{(1)}_{ij}$ is a reflection that this mode remains unconstrained. This is properly interpreted as a new holographic source, whose dual operator is the partially massless response function, introduced in Ref.~\cite{Grumiller:2013mxa, Deser:2012qg}.

In turn, Eq.~\eqref{delta g} leads to 
\begin{equation}
    \delta {g}_{ij}^{(2)} =   \bar{D}_{i}^{(0)} \xi^{(2)}_{j} + \bar{D}_{j}^{(0)} \xi^{(2)}_{i} \,  \label{g2} \,,
\end{equation}
whose solution shows that $g^{(2)}_{ij}$ features a universal value in terms of the Schouten tensor of the holographic metric ~\cite{Imbimbo:1999bj} 
\begin{equation}
    g^{(2)}_{ij} = - \frac{\ell^2}{d-1}\left( \mathscr{R}^{(0)}_{ij} - \frac{1}{2\left(d-2\right)} \mathscr{R}^{(0)} g^{(0)}_{ij} \right) \, ,
\end{equation}
even in the presence of a non-vanishing source $g^{(1)}_{ij}$.
In an analogous fashion, one may partially determine the form of other coefficients in FG expansion.

The kinematic study above is a requirement for the asymptotic description of surfaces embedded in AAdS ambient space and anchored to its conformal boundary~\cite{schwimmer2008entanglement, Hung:2011xb}.

\subsection{A Fefferman--Graham--like expansion for submanifolds}

Consider a codimension-2 submanifold $\Sigma$ embedded in $\mathcal{M}$, with its boundary $\partial \Sigma$ anchored to the conformal boundary $\partial \mathcal{M}$.
Let $Y^{a} = (\tau, y^u)$ be the local coordinates in $\Sigma$ and $X^{\mu} = (z, x^{i})$ the coordinates on the bulk $\mathcal{M}$, where the embedding functions $X^{\mu} = X^{\mu} (Y^{a})$ induced the metric on $\Sigma$ by

\begin{equation}
    \gamma_{ab} =  \frac{\partial X^{\mu}}{\partial Y^a} \frac{\partial X^{\nu}}{\partial Y^b} g_{\mu \nu}  \, . \label{gamma}
\end{equation}
The reparametrization invariance on $\Sigma$ is fixed by the gauge conditions

\begin{equation}
 \tau = z  \qquad \text{and} \qquad \gamma_{u \tau} = 0 \, , \label{gaugePBHsigma}
\end{equation}
reducing the intrinsic metric of the surface to the Gauss normal form

\begin{equation}
    ds_{\gamma}^{2} =   \gamma_{z z} \, d{z}^{2} + \bar{\gamma}_{uv} \, dy^{u}  dy^{v} \, . \label{induced metric}
\end{equation}
Assuming that the embedding functions $x^{i}(z, y)$ admits a near-boundary expansion of the form

\begin{align}
    x^{i}(z, y) = \, x^{(0)i}(y) + z \,x^{(1)i}(y) + z^{2} x^{(2)i}(y) + \mathcal{O}\left( z^{3}\right) \, , \label{expansionxi}
\end{align}
--where $x^{(0)i}(y)$ describes the embedding of $\partial \Sigma$ on the conformal boundary $\partial \mathcal{M}$-- the metric  $\bar{g}_{ij}$ is expanded as

\begin{equation}
\bar{g}_{ij}(z, y) = g^{(0)}_{ij} (y) + z  \, \mathcal{G}^{(1)}_{ij}(y)  + z^2 \, \mathcal{G}^{(2)}_{ij}(y)  + \mathcal{O}\left( z^3 \right) \, . \label{expansion of g(z,y)}
\end{equation}
Here, the next-to-lead order of the above metric is given  in terms of non-Einstein modes as

\begin{equation}
\mathcal{G}^{(1)}_{ij}(y) = \, x^{(1)k} \partial_k g^{(0)}_{ij} + g^{(1)}_{ij} \, .
\end{equation}
Using Eqs. \eqref{gamma} and \eqref{expansionxi}, the $\gamma_{zz}$ component of the induced metric takes the form

\begin{equation}\label{gammazzexpansion}
     \gamma_{zz} =   \, \frac{\partial X^{\mu}}{\partial z} \frac{\partial X^{\nu}}{\partial z} {g}_{\mu \nu} = \frac{1}{z^{2}} \left( N^{(0)}(y) + z \,  N^{(1)}(y) + z^{2} \, N^{(2)}(y) + \mathcal{O}(z^3) \right) \, ,  
\end{equation}
where the coefficients $N^{(0)}(y)$ and $N^{(1)}(y)$ are determined by the relations

\begin{align}
N^{(0)}(y) = &{} \, \ell^2 +  x^{(1)i} x^{(1)j} g^{(0)}_{ij} \, , \\
N^{(1)}(y) = &{} \,  x^{(1)i} x^{(1)j} \mathcal{G}^{(1)}_{ij} 
+ 2 \left( x^{(1)i} x^{(2)j} + x^{(2)i} x^{(1)j} \right) g^{(0)}_{ij}  \, .
\end{align}
In a similar fashion, the boundary components of the surface metric are  expressed as

\begin{equation}
    \bar\gamma_{uv}=  \frac{\partial X^{\mu}}{\partial y^u} \frac{\partial X^{\nu}}{\partial y^v} g_{\mu \nu}  = \frac{1}{z^2} \sigma_{uv}  \, , \label{eq:modesgamma}
\end{equation}
where $\sigma_{uv}$ is given in terms of the power-series expansion

\begin{equation}
    \sigma_{uv}(z, y) = \sigma_{uv}^{(0)}(y) + z \, \sigma_{uv}^{(1)}(y) + \, z^2 \sigma_{uv}^{(2)}(y) + \mathcal{O}(z^3) \, .
\end{equation}
In order to work out the holographic properties of $\partial \Sigma$, one identifies the leading-order term with the conformal metric on it as

\begin{equation}
\sigma^{(0)}_{uv}(y) = \partial_u x^{(0)i} \partial_v x^{(0)j} g^{(0)}_{ij} \, ,
\end{equation}
while the linear order in $z$ is written as

\begin{align}
\sigma^{(1)}_{uv}(y) = &{} \Bigl( \partial_u x^{(1)i} \partial_v x^{(0)j} + \partial_u x^{(0)i} \partial_v x^{(1)j} \Bigr) g^{(0)}_{ij}
+ \partial_u x^{(0)i} \partial_v x^{(0)j} \mathcal{G}^{(1)}_{ij}  \, .
\end{align}
In principle, Eq.\eqref{eq:modesgamma},  determines the structure of the higher-order modes. However, these terms will not be relevant for the analysis below.

The vanishing of the crossed components of the induced metric 

\begin{equation}
    \gamma_{zu} =  \, \frac{\partial X^{\mu}}{\partial z} \frac{\partial X^{\nu}}{\partial y^u} {g}_{\mu \nu}  \, ,
\end{equation}
when expanded using Eqs. \eqref{expansionxi} and \eqref{expansion of g(z,y)}, at the leading order, gives the constraint

\begin{equation}
    x^{(1)i} \partial_{u} x^{(0)j} g_{ij}^{(0)} = 0 \, .
\end{equation}
This relation implies that $x^{(1)i}$ is orthogonal to the tangential directions of $\partial \Sigma$, yielding

\begin{equation}
    x^{(1)i} = \left| x^{(1)} \right| n^{i} \, , \label{x^(1)i}
\end{equation}
where $n^{i}$ is the unit vector to $\partial \Sigma$ in $\partial \mathcal{M}$, and  $\left| x^{(1)} \right| = \sqrt{x^{(1)i} x^{(1)j} g^{(0)}_{ij}}$. 


Generic PBH transformations in the bulk may well violate the gauge conditions~\eqref{gaugePBHsigma} on the surface.
In order to restore these conditions, it is necessary to introduce the compensating diffeomorphisms

\begin{equation}
   \tau' = \tau + \tilde{\xi}^{\tau} (\tau,y) \, , \qquad y'^{u} = y^{u} +  \tilde{\xi}^{u} (\tau,y) \, , 
\end{equation}
constrained by $\delta z = \delta \tau$ and $\delta \gamma_{a \tau} = 0$. 
These restrictions imply an asymptotic behavior

\begin{align}
    \tilde{\xi}^{z}(z, y) = &{} - z \, \Omega\left( x \right)  \, , \label{xitau}\\ 
    \tilde{\xi}^{u}(z, y) = &{}  \int^{z}_{0}  z' \gamma_{zz}(z', y) \gamma^{uv}(z', y) \ \partial_{v} \Omega\left( x \right) \ dz' \label{eq:xiu}\, .
\end{align}
A more explicit form of the above equations show that these terms can be expanded in powers of $z$ as

\begin{align}
       \tilde{\xi}^{z}(z, y) &{} =  - z \, \Omega^{(0)} - z^{2} \left( x^{(1)i} \, \partial_{i} \Omega^{(0)} \right) + \mathcal{O}\left( z^3 \right) \, , \\
       \tilde{\xi}^{u}(z, y) &{} = \frac{1}{2} z^2 \left( \ell^2 + x^{(1)i} \,x^{(1)j} \, g_{ij}^{(0)} \right) \sigma^{(0)uv} \partial_{v} \Omega^{(0)} + \mathcal{O}\left( z^3 \right) \, ,
\end{align}
where $\Omega^{(0)}$ is a function $\Omega^{(0)} \equiv \Omega(x^{(0)})  $. 
As it is pointed out in Ref.\cite{Chu:2017aab}, the variation of the embedding functions $x^{i}(z,y)$ under PBH transformations reveals the universal nature of the modes $x^{(1)i}$ and $x^{(2)i}$. 
Such variation is given by

\begin{equation}
    \delta x^{i}(z, y) = \tilde{\xi}^{a}(z, y) \, \partial_{a} x^{i}(z, y) - \xi^{i}\left(x^{i}(z, y)\right) \, .
\end{equation}
At leading order, the boundary embedding remains invariant

\begin{equation}
\delta x^{(0)i} = 0 \, ,
\end{equation}
while, at first order in $z$, the mode $x^{(1)i}$ undergoes a Weyl rescaling 

\begin{equation}
\delta x^{(1)i} = - \, \Omega^{(0)} x^{(1)i} \, ,
\end{equation}
in an analogous way as to the transformation of $g^{(1)}_{ij}$. This result indicates the universality of $x^{(1)i}$, in the sense that the surface diffeomorphisms preserve $x^{(1)i}$ along the normal direction.

At quadratic order in $z$, the variation of $x^{(2)i}$ is determined by the formula

\begin{align}
\delta x^{(2)i} \;=\; &{}
-\,2\,\Omega\,x^{(2)i}
\;+\;\frac{1}{2}\,\lvert x^{(1)}\rvert^{2}\, \sigma^{(0)uv} \partial_{u} x^{(0)i}  \partial_{v} x^{(0)j} \,\partial_{j}\Omega \nonumber \\
&{} \quad\;-\;\frac{1}{2}\,\left(\ell^2\;+\;2\,\lvert x^{(1)}\rvert^{2}\right)\,n^{i}\,n^{j}\,\partial_{j}\Omega
\,.
\end{align}
The proper use of the completeness relation at leading order, i.e.,

\begin{equation}
  \sigma^{(0)uv} \partial_{u} x^{(0)i}  \partial_{v} x^{(0)j} = g^{(0)ij} - n^{i} n^{j}\,,
\end{equation}
together with Eq.\eqref{x^(1)i}, leads to  the uniquely fixed mode $x^{(2)i}$, that reads

\begin{equation}
    x^{(2)i} = \frac{\ell^{2} + \left| x^{(1)} \right|^2}{4} \kappa^{i} - \frac{1}{4} \sigma^{(0)uv} \partial_{u}x^{i} \partial_{v} x^{j} \partial_{j} \left| x^{(1)} \right|^2 - \frac{1}{2} \Gamma^{(0)i}_{kl} n^{k} n^{l} \left| x^{(1)} \right|^2  \, ,
\end{equation}
as shown in Ref.\cite{Chu:2017aab}. Here, $\kappa^i=\sigma^{(0)uv} \kappa^{i}_{uv}$ stands for the trace of the extrinsic curvature of $\partial \Sigma$ given by

\begin{equation}
    \kappa^{i}_{uv} =  
\partial_u \partial_v x^{(0)\,i} 
- \zeta_{uv}^{(0)w} \partial_w x^{(0)\,i} 
+ \Gamma^{(0)\,i}_{\quad jk} \partial_u x^{(0)\,j} \partial_v x^{(0)\,k} \,,
\end{equation}
where $\Gamma^{(0)i}_{jk}$ and $\zeta_{uv}^{(0)w}$ are the Christoffel symbols of $g^{(0)}_{ij}$ and $\sigma^{(0)}_{ij}$, respectively. This result extends the universal character of $x^{(2)i}$ found in Ref.\cite{Chu:2017aab} for a non-Einstein ambient spacetime, i.e., when the $g_{(1) ij}$ mode is switched on..

In sum, the analysis presented makes explicit the asymptotic behavior of the induced metric on $\Sigma$. Equipped with this result, in what follows, the functional codimension--two $L_{\Sigma}$ is proved to be finite to any surface embedded in an AAdS manifold $\mathcal{M}$.

\subsection{Finiteness of the conformal codimension-2 functionals}

The conical contribution when CG is considered as a gravitational theory in the bulk is

\begin{align}
L_{\Sigma} = \beta \int\limits_{\Sigma} d^2 Y \sqrt{\gamma} \left( W^{ab}_{ab} - P^{A}_{ab} P_{A}^{ab} \right) - 4 \pi \beta \chi \left(\Sigma \right) \, . \label{Lconf}
\end{align}
It emerges as a Weyl-invariant functional induced from the action by the LM prescription in the context of QCG.
Both terms under the integral sign in the above formula are conformal invariants. However, CG makes them appear in a particular combination known in the literature as Graham--Witten anomaly~\cite{Graham:1999pm}. This surface functional recovers the renormalized area and reduced Hawking mass for an Einstein ambient space and different conditions on $\Sigma$.  At the same time, for a pure AdS background spacetime, $L_{\Sigma}$ turns into the Willmore energy of a closed surface \cite{anastasiou2022energy}.

The relation between Willmore energy and renormalized Holographic Entanglement Entropy for vacuum states was emphasized by Fonda et al. in Ref.~\cite{Fonda:2015nma}. It is of particular relevance the existence of a topological bound in Willmore energy, that extremizes the entanglement entropy for disk configurations in three-dimensional CFTs. This claim was later generalized for scalar and fermionic fields in Ref.~\cite{Bueno:2021fxb}.

In order to work out the finiteness of the functional ${L}_{\Sigma}$, one should make explicit the asymptotic  form of the terms $P_{ab}P^{ab}$ and $W_{ab}^{ab}$. 
It is important to first examine the embedding of the boundary $\partial \Sigma$ in $\mathcal{M}$, as this would allow to characterize the leading-order contributions to the functional of interest. 
The analysis is performed for quantities carrying spacetime indices, as a consequence of the use of orthonormal bases and normal vectors, i.e., $T^{\mu}_{\rho \sigma} = T^{A}_{ab} n^{\mu}_{A} e^{a}_{\rho} e^{b}_{\sigma}$. 

\begin{figure}[H]
\centering
\includegraphics[width=0.75\textwidth]{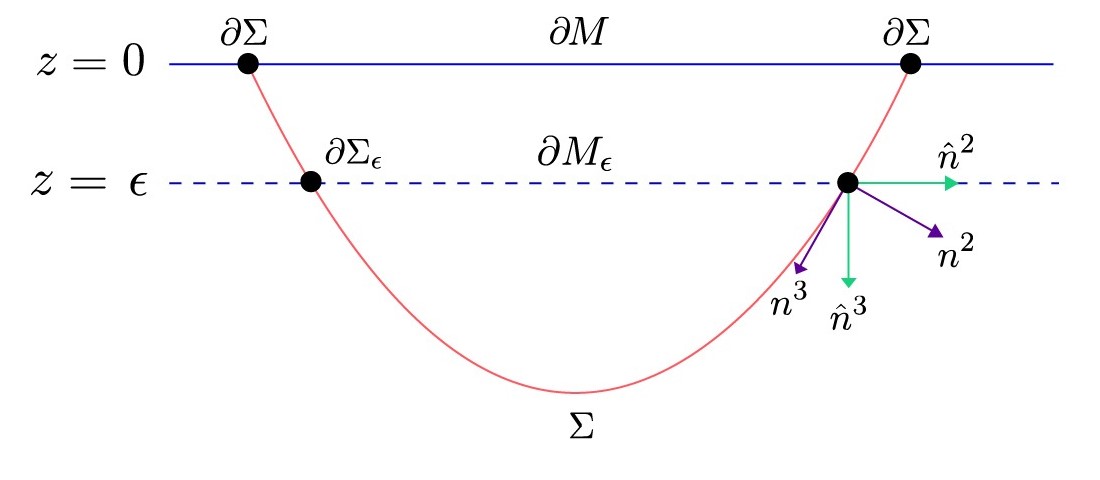}
\caption{
The boundary $\partial\Sigma$ lies at $z=0$. The two sets of normal vectors, $\bigl(n^2, n^3\bigr)$ and $\bigl(\hat{n}_2, \hat{n}_3\bigr)$ provide  two equivalent characterizations of $\partial\Sigma$. The time-directed vector $n^1$ is omitted for clarity. 
\label{fig:relation between normal vectors}}
\end{figure}

As depicted in Fig.~\ref{fig:relation between normal vectors}, one may view the boundary $\partial \Sigma$ embedded in $\mathcal{M}$ in two equivalent ways. First, from the perspective of $\Sigma$ itself, introduce the orthonormal set of vectors $\left( n^{1}, n^{2}, n^{3} \right)$ where $n^1$ is aligned with the time direction, $n^2$ is normal to $\Sigma$ into $\mathcal{M}$, whereas $n^{3}$ is an inward--pointing vector tangent to $\Sigma$.  

Alternatively, considering the FG gauge of the bulk metric, one may take the normal vectors $\left( {n}_{1}, \hat{n}_{2}, \hat{n}_{3} \right)$, where $\hat{n}_{2}$ is tangent to the boundary $\partial \mathcal{M}$, and $\hat{n}_{3}$ points along the $z$--direction.

The bulk metric is described by the orthonormal triad $\left( {n}_{1}, \hat{n}_{2}, \hat{n}_{3} \right)$, which generates the bulk metric in the FG gauge \eqref{FG gauge}, duly expressed as

\begin{equation}
    ds^{2} = \frac{\ell^2}{z^2} dz^2 + \frac{1}{z^2} \left( dt^2 + \alpha{(z,x)} dr^2 + \bar{\gamma}_{uv} dy^u dy^v \right)  \, ,
\end{equation}
where the function $\alpha{(z,x)}$ characterizes the geometry along the normal $\hat{n}^{2}_{\mu}$ and $\bar{\gamma}_{u v}$ is the induced metric of $\partial {\Sigma}$ given in Eq.~\eqref{eq:modesgamma}.
In this coordinate frame, the components of these normal vectors are given by

\begin{align}
    n^{1}_{\mu} = \frac{1}{z} \delta_{\mu}^{t} , \quad  \hat{n}^{2}_{\mu} = \frac{\alpha{(z,x)}}{z} \delta^{r}_{\mu}  , \quad  \hat{n}^{3}_{\mu} = \frac{\ell}{z} \delta_{\mu}^{z} \, . \label{contravariant normal vectors}
\end{align}
Equipped with these normal vectors, the corresponding extrinsic curvatures of $\partial \Sigma$ are computed as Lie derivatives

\begin{align}
    \kappa_{\mu \nu} {}& = \frac{1}{2} \mathcal{L}_{\hat{n}^2} \bar{\gamma}_{\mu \nu} =
     \frac{1}{2 z \alpha} \left(  \partial_{r} \sigma_{\mu \nu}  - \frac{2}{\alpha} {\sigma}_{r ( \mu} \partial_{\nu )} \alpha  \right) \, , \\
    \kappa^{\perp}_{\mu \nu}
  {}&  = \frac{1}{2} \mathcal{L}_{\hat{n}^3}  \bar{\gamma}_{\mu \nu}  = \frac{1} {\ell} \left( - \frac{\sigma_{\mu \nu}}{z^2} + \frac{\partial_{z} {\sigma}_{\mu \nu}}{2 z} \right)  \, ,
\end{align}

whose traces are given by
\begin{align}
    \kappa = {}& \frac{z}{2\alpha} \left( \sigma^{\mu \nu} \partial_{r} \sigma_{\mu \nu} - \frac{2}{\alpha} \partial_{r} \alpha \right) \, , \\
    \kappa^{\perp} = {}&  \frac{1}{\ell} \left( - 1 + \frac{z}{2} \sigma^{\mu \nu} \partial_{z} \sigma_{\mu \nu} \right) \, .
\end{align}
At each surface $\partial \Sigma_{\epsilon}$ located at $z=\epsilon$ (see Fig. \ref{fig:relation between normal vectors}), the set of vectors $(n_2, n_3)$ and $(\hat{n}_{2}, \hat{n}_{3})$ can be connected through the transformations

\begin{align}
    {n}^2 ={}& A{(z, x)} \hat{n}^{2}  + A^{\perp}{(z, x)} \hat{n}^{3} \, , \label{n2-hn2hn3} \\
    {n}^3 ={}& - A^{\perp}{(z, x)} \hat{n}^{2} + A{(z, x)} \hat{n}^{3} \, , \label{n3-hn2hn3}
\end{align}
where the coefficients satisfy the orthogonality condition

\begin{equation}
A^2{(z, x)}  + {A^{\perp}}{(z, x)} ^2 = 1 \, .    
\end{equation}
The fact $\Sigma$ is orthogonally anchored to $\partial \mathcal{M}$, is translated into the boundary conditions

\begin{equation}
  A^{\perp}{(0, x)} = 0 \, , \qquad  A{(0, x)} = 1 \, ,
\end{equation}
which imply that the normal vectors $(\hat{n}^2, \hat{n}^3)$ coincide with $(n^2, n^3)$ at $z=0$.

At this point, one needs to perform the asymptotic analysis for the behavior of the co-dimension 2 conformal invariants which enter in the energy functional \eqref{Lconf}. For the the term involving the traceless part of the extrinsic curvature, i.e.,

\begin{equation} \label{P2}
P_{\mu \nu} P^{\mu \nu}= \mathcal{K}^{A}_{\mu \nu}\mathcal{K}_{A}^{\mu \nu} - \frac{1}{2} \mathcal{K}^{A} \mathcal{K}_{A} \, ,
\end{equation}
one may notice that, in the static case, $\mathcal{K}_{\mu \nu}^{(1)} = 0$. Hence, one may denote $\mathcal{K}_{\mu \nu}^{(2)} = \mathcal{K}_{\mu \nu}$ for simplicity, that is cast in the form

\begin{equation}
    \mathcal{K}_{\mu \nu} = \gamma^{\alpha}_{\mu} \gamma^{\beta}_{\nu} \nabla_{\alpha} n^{2}_{\beta} \, , \label{Extrinsic curvature of Sigma}
\end{equation}
while replacing $\mathcal{K}^{A} \mathcal{K}_{A}=\mathcal{K}^2$.
Upon substitution of Eqs. \eqref{contravariant normal vectors} and \eqref{n2-hn2hn3} into the definition of the extrinsic curvature $\mathcal{K}_{\mu \nu}$, produces

\begin{equation}
\mathcal{K}_{\mu \nu} = A\kappa_{\mu \nu} + A^{\perp} \kappa_{\mu \nu}^{\perp} 
    + \hat{n}_{\mu}^{3} \hat{n}_{\nu}^{3} U 
    \, , \label{Kmunu}
\end{equation}
where the function $U$ reads

\begin{align}
    U & \equiv \frac{z}{\alpha} \partial_{r} A  + \frac{z}{\ell} \partial_{z} A^{\perp} - \frac{1}{\ell}\left(1 - z \frac{\partial_{z} \alpha}{\alpha} \right) A^{\perp} \, . \label{U} 
\end{align}
Considering a power-series expansion in $z$ for the functions $\alpha{(z, x)}$, $A{(z, x)}$, and $A^{\perp}{(z, x)} $, of the form

\begin{align}
    \alpha(z, x) = {}& \sum^{\infty}_{n=0} z^{n} \alpha^{(n)}(x) \, , \\
    \qquad A(z, x) = {}& \sum^{\infty}_{n=0} z^{n} A^{(n)}(x) \, , \\ 
   \qquad A^{\perp}(z, x) = {}& \sum^{\infty}_{n=0} z^{n} A^{\perp(n)}(x) \, ,
\end{align}
leads to an expression for the square of the trace of the extrinsic curvature 
up to quadratic order in $z$

\begin{equation}
\mathcal{K}^2= \left[\mathcal{K}^2 \right]^{(0)} +z \left[\mathcal{K}^2 \right]^{(1)}+\mathcal{O}(z^2)\,,
\label{Ksquared}
\end{equation}
where

\begin{align}
\left[\mathcal{K}^2 \right]^{(0)}
\;=\;&
\frac{4\bigl(A^{\perp(0)}\bigr)^2}{\ell^2} \, , \label{K0} \\
\left[\mathcal{K}^2 \right]^{(1)}
\;=\;&
-\;\frac{2\,A^{\perp(0)}\,A^{(0)}}{\ell\,\alpha^{(0)}}
\Bigl[
    \sigma^{(0)\mu\nu}\,\partial_r \sigma^{(0)}_{\mu\nu}
    \;-\;
    \,\frac{2}{\alpha^{(0)}}\,\partial_r \alpha^{(0)}
\Bigr]
\;-\;
\frac{2\,\bigl(A^{\perp(0)}\bigr)^2}{\ell^2}
\,\sigma^{(1)} \nonumber
\\
&
+\;\frac{4\,A^{\perp(0)}\,A^{\perp(1)}}{\ell^2}
\;-\;
\frac{4\,A^{\perp(0)}}{\ell\,\alpha^{(0)}}
\,\partial_r A^{(0)}
\;-\;
\frac{4\,\bigl(A^{\perp(0)}\bigr)^2\,\alpha^{(1)}}
      {\ell^2\,\alpha^{(0)}} \, . \label{K2}
\end{align}
In a similar fashion, the square of the extrinsic curvature tensor 
admits the expansion

\begin{equation}
\mathcal{K}_{\mu \nu} \mathcal{K}^{\mu \nu} = \left[ \mathcal{K}_{\mu \nu} \mathcal{K}^{\mu \nu} \right]^{(0)} + z \left[ \mathcal{K}_{\mu \nu} \mathcal{K}^{\mu \nu} \right]^{(1)} + \mathcal{O}(z^2) \,,
\label{Kmunusquared}
\end{equation}
where

\begin{align}
\left[ \mathcal{K}_{\mu \nu} \mathcal{K}^{\mu \nu} \right]^{(0)} \;=\;& \frac{2 (A^{\perp(0)})^2}{\ell^2} \, , \label{KK^0} \\
\left[ \mathcal{K}_{\mu \nu} \mathcal{K}^{\mu \nu} \right]^{(1)}\;=\;&
-\;\frac{\,A^{\perp(0)}\,A^{(0)}}{\ell\,\alpha^{(0)}}
\Bigl[
    \sigma^{(0)\mu\nu}\,\partial_r \sigma^{(0)}_{\mu\nu}
    \;-\;
    \,\frac{2}{\alpha^{(0)}}\,\partial_r \alpha^{(0)}
\Bigr]
\;-\;
\frac{\bigl(A^{\perp(0)}\bigr)^2}{\ell^2}
\,\sigma^{(1)} \nonumber
\\
&
+\;\frac{2\,A^{\perp(0)}\,A^{\perp(1)}}{\ell^2}
\;-\;
\frac{2\,A^{\perp(0)}}{\ell\,\alpha^{(0)}}
\,\partial_r A^{(0)}
\;-\;
\frac{2\,\bigl(A^{\perp(0)}\bigr)^2\,\alpha^{(1)}}
      {\ell^2\,\alpha^{(0)}} \, . \label{KK^1}
\end{align}
Combining Eqs.~\eqref{Ksquared} and \eqref{Kmunusquared}, the expression for $P_{\mu \nu} P^{\mu \nu}$ reads

\begin{equation}
    P_{\mu \nu} P^{\mu \nu} = \left( \left[ \mathcal{K}_{\mu \nu} \mathcal{K}^{\mu \nu} \right]^{(0)} -  \frac{1}{2}\left[\mathcal{K}^2 \right]^{(0)} \right) + z \left( \left[ \mathcal{K}_{\mu \nu} \mathcal{K}^{\mu \nu} \right]^{(1)} -  \frac{1}{2}\left[\mathcal{K}^2 \right]^{(1)}\right) + \mathcal{O}\left( z^2 \right) \, .
\end{equation}
Notice the cancelation of the first two order contributions, that induces a quadratic order in $z$ fall-off for this term of $L_{\Sigma}$.

At this level, one needs to work out the asymptotic behavior of other term, $W^{ab}_{ab}$, which enters in the co-dimension 2 conformal functional. As established in Sec.~\ref{section 2}, the bulk Weyl tensor exhibits an asymptotic fall--off of order $z^2$ or higher, what implies that its projection onto the surface $\Sigma$ produces the same effect on the surface functional. As a matter of fact,

\begin{equation}
    W_{ab}^{ab}  = \mathcal{O}\left( z^2 \right) \, .
\end{equation}
From the Fefferman--Graham like form of the induced metric on the surface, one can obtain the explicit expansion of $\sqrt{\gamma}$ , given by the expression

\begin{align}
    \sqrt{\gamma} = \frac{\sqrt{\sigma^{(0)}}}{z^2}\left[ {N^{(0)}}^{1/2} + \frac{z}{2} \left( {N^{(0)}}^{1/2} \sigma^{(1)} + \frac{N^{(1)}}{{N^{(0)}}^{1/2}} \right) +  \mathcal{O}\left(z^2\right) \right] \, .
\end{align}
As a consequence, the conformal functional $L_{\Sigma}$ is proved to be

\begin{equation}
    L_{\Sigma} \sim \int\limits_{\Sigma} d^2 Y \frac{\sqrt{{\sigma}^{(0)}}}{z^{2}}\mathcal{O} \left( z^2 \right) = \text{finite} + \mathcal{O} \left( z \right)\, .
\end{equation}
Hence, this surface functional is finite for arbitrary codimension--$2$ surfaces (i.e., minimal or not) embedded in AAdS manifolds in four dimensions. This fact ensures the finiteness of the reduced Hawking mass, renormalized area and Willmore energy under the corresponding assumptions.

\section{Conclusions} \label{Conclusions}
Energy functionals play an essential role in the description of surface properties in areas as diverse as gravitational physics and biology \cite{Lomholt_2006,2017arXiv170904399C,isenberg1978science,reilly1982mean,penrose1973naked}. In this work, the functional $L_{\Sigma}$ is singled out among all possible codimension--$2$ structures which may be derived from QCG, by demanding conformal invariance on the surface. Indeed, by applying the LM prescription to QCG, one may prove that CG is the only gravity theory in the bulk where the conical contribution is a combination of Weyl invariants.

In the same vein, the finiteness of $L_{\Sigma}$, was rendered manifest by the use of a FG-like expansion for submanifolds under relaxed AAdS asymptotics, even for non-minimal surfaces intersecting the conformal boundary at an arbitrary angle. In fact, the inclusion of higher-derivative modes of CG in the bulk FG gauge has direct consequences on the embedding of surfaces. The detailed analysis of the induced metric unveils the cancellation of infrared divergences  up to the relevant order. In this expansion,  both terms $P_{ab} P^{ab}$ and $W^{ab}_{ab}$ decay as $\mathcal{O}\left(z^2 \right)$, what guarantees that $L_{\Sigma}$ remains finite without the need of counterterms. This result provides yet another example of the connection between conformal invariance and renormalization, a defining feature of the Conformal Renormalization scheme for AdS gravity. This feature is not limited to four dimensions. Actually, recent results \cite{Anastasiou:2024rxe} show the four-dimensional analogue of $L_{\Sigma}$, dubbed Graham-Reichert energy, as coming from the CG theory which has an Einstein sector in six dimensions. This functional  gives rise to a renormalized holographic entanglement entropy for a wide variety of surfaces.

The action of Einstein-AdS gravity is on-shell proportional to the volume of the spacetime. The removal of infinities, coming from the conformal boundary of AAdS spaces, leading to the notion of Renormalized Volume, has been linked to conformal invariants in mathematical literature \cite{Anderson2000L2CA,Chang:2005ska}. The results shown here are a manifestation of the fact Renormalized Volume induces Renormalized Area, and other closely related finite functionals in codimension-2.

\section*{Acknowledgements}
We thank Ignacio J. Araya, Nicolas Boulanger and Javier Moreno for useful discussions. The work of GA is funded by ANID FONDECYT grants No. 11240059 and 1240043. RO was supported by Anillo Grant ACT210100 \emph{Holography and its applications to High Energy Physics, Quantum Gravity and Condensed Matter Systems} and ANID FONDECYT Regular grants No. 1230492, 1231779, 1240043 and 1240955. MB is supported by Becas de Magister UNAP, ANID FONDECYT grants No. 11240059 and 1230492 and Anillo Grant ACT210100.

\appendix
\section{Notation and Conventions}
This section outlines the notation and conventions used throughout this work. The bulk manifold $\mathcal{M}$ is a spacetime equipped with the metric $g_{\mu \nu}$, whose boundary is $\partial \mathcal{M}$. The surface $\Sigma$ lives in codimension--$2$, and it is embedded in $\mathcal{M}$ by two normal vectors. Its boundary is given by $\partial \Sigma$. Table~\ref{table:notation} summarizes the notation used. 
 
\begin{table}[h!]
\centering
\renewcommand{\arraystretch}{1.5} 
\begin{tabular}{|c|c|c|c|c|}
\hline
\multicolumn{1}{|c|}{} & $\mathcal{M}$ & $\partial \mathcal{M} \subset \mathcal{M}$ & $\Sigma \subset \mathcal{M}$ & $\partial \Sigma \subset \Sigma$ \\ \hline
Indices & $\mu,\nu, \ldots$ & $i,j, \ldots$ & $a,b, \ldots$ & $u,v, \ldots$ \\ 
Metric & $g_{\mu \nu}$ & $h_{ij}$ & $\gamma_{ab}$ & $\bar{\gamma}_{uv}$ \\ 
Riemann tensor & $R_{\mu \nu \gamma \delta}$ & $\mathscr{R}_{ijkl}$ & $\mathcal{R}_{abcd}$ &  \\ 
Extrinsic curvature &  & $\mathscr{K}_{ij}$ & $\mathcal{K}_{ab}$ & $\kappa_{uv}$ \\ \hline
\end{tabular}
\caption{Notation for the spacetime and corresponding submanifolds}
\label{table:notation}
\end{table}

The embedding of $\Sigma$ in $\mathcal{M}$ is given by the functions $ X^{\mu} = X^{\mu} \left( Y^{a} \right) $, where $X^{\mu}$ are the coordinates of the bulk and $Y^a$ are the intrinsic coordinates on  $\Sigma$.
The induced metric on $\Sigma$ is constructed by the pullback of the bulk metric,
\begin{equation}
    \gamma_{ab} = e^{\mu}_{a} e^{\nu}_{b} g_{\mu \nu}
    \label{eq:metricasigma}
\end{equation}
where $e^{\mu}_{a} = \partial_{a}X^{\mu}$  forms a basis of the tangent space of $\Sigma$.  
The normal bundle of the surface is described by the orthonormal vectors $n^{\mu}_{A}$, which are also orthogonal to each other. 
The metric of the normal bundle is  defined as, 
\begin{equation}
    \delta_{AB} = n^{\mu}_{A} n^{\nu}_{B} g_{\mu \nu}\,,
    \label{eq:metricanormal}
\end{equation}
where the indices $A$, $B$, $\dots$ represent the normal directions to $\Sigma$. The completeness relation between the tangential and normal directions is given by 
\begin{equation}
    g^{\mu \nu} = \gamma^{ab} e^{\mu}_{a} e^{\nu}_{b} + \delta^{AB} n^{\mu}_{A} n^{\nu}_{B} \, .
    \label{eq:completerelations}
\end{equation}
where $\gamma^{ab}$ and $\delta^{AB}$ are the inverse of the tangent and normal parts of the metric tensor, respectively. 

The extrinsic curvature $\mathcal{K}^{A}_{ab}$ of the surface along the normal vector $n^{\mu}_{A}$ is defined as
\begin{align}
    \mathcal{K}^{A}_{ab} \equiv e^{\mu}_{a} e^{\nu}_{b} {\nabla}_{\mu} n^{A}_{\mu} \, . 
\end{align}
In order to work with a covariant derivative which acts on tensors of mixed type --either in the spacetime or on $\Sigma$-- one may consider the Van der Waerden--Bortolotti covariant derivative. Its action on a tensor $T^{\mu}_{j A}$ is given by 
\begin{equation}
    \nabla_{a} T^{\mu}_{b A} = \partial_{a} T^{\mu}_{b A} + {\Gamma}^{\mu}_{\sigma \rho} e^{\sigma}_{a} T^{\rho}_{b A} - \Gamma^{c}_{ab} T^{\mu}_{c A} - \Gamma^{B}_{a A}  T^{\mu}_{b B} \, ,
\end{equation}
where $\Gamma^{c}_{ab}$ is the Christoffel symbols associated to the metric $\gamma_{ab}$. The relation to the Christoffel symbol of the bulk, ${\Gamma}^{\mu}_{\sigma \rho}$, is given by
\begin{align}
    \Gamma^{c}_{ab} ={}& \left( \partial_{a} e^{\mu}_{b} + {\Gamma}^{\mu}_{\sigma \nu} e^{\sigma}_{a} e^{\nu}_{b} \right) e^{c}_{\mu}\,, \\
    \Gamma^{A}_{Ba} ={}& \left( \partial_{a} n^{\mu}_{B} + {\Gamma}^{\mu}_{\sigma \nu} e^{\sigma}_{a} n^{\nu}_{B} \right) n^{A}_{\mu}\,.
\end{align}
In this notation, the Gauss--Weingarten equations are expressed as
\begin{align}
    \nabla_{a} e^{\mu}_{b} ={}& - \mathcal{K}^{A}_{ab} n^{\mu}_{A} \, , \label{eq:covariantofe}\\
    \nabla_{a} n^{\mu}_{B} ={}& \mathcal{K}^{b}_{B a} e^{\mu}_{b} \, .
    \label{eq:covariantofn}
\end{align}
The relation between the Riemann tensor of the bulk with the one of $\Sigma$ is governed by the Gauss relations, which state that 
\begin{equation}
e_{k}^{\mu }e_{a}^{\nu }e_{d}^{\rho }e_{b}^{\sigma }R_{\mu \nu \rho \sigma } =\mathcal{R}_{cadb} +\left(\mathcal{K}_{cb}^{A}\mathcal{K}_{A ad} -\mathcal{K}_{cd}^{A}\mathcal{K}_{A ab} \right) \, , \label{GaussRel}
\end{equation}
\begin{equation}
e_{a}^{\nu }e_{b}^{\sigma }R_{\nu \sigma } -R_{\mu \nu \rho \sigma }e_{a}^{\nu }e_{b}^{\sigma }n_{B}^{\mu }n^{\rho B} = \mathcal{R} _{ab} +\left(\mathcal{K}_{b}^{c A}\mathcal{K}_{ac A} -\mathcal{K}^{A}\mathcal{K}_{A ab}\right) \, , \label{GaussCont}
\end{equation}
\begin{equation}\label{GaussScalarRelation}
R -2R_{\nu \rho }n_{B}^{\mu }n^{\sigma B} +R_{\mu \nu \rho \sigma }n_{B}^{\nu }n^{\sigma B}n_{A}^{\mu }n^{\rho A} = \mathcal{R}  +\left(\mathcal{K}_{ab}^{A}\mathcal{K}_{A}^{ab} -\mathcal{K}^{A}\mathcal{K}_{A}\right) \,.
\end{equation}
where $\mathcal{R}_{abcd}$ is the Riemann tensor associated to the induced metric $\gamma_{ab}$.
The remaining projections of the curvature tensors of the bulk into $\Sigma$ are given by the Codazzi--Mainardi relations,
\begin{equation}
R_{\mu \nu \rho \sigma }e_{c}^{\mu }e_{a}^{\nu }e_{b}^{\sigma }n_{A}^{\rho } = \nabla _{c}\mathcal{K}_{Aab} - \nabla_{a}\mathcal{K}_{A bc} \, , \label{codazzi}
\end{equation}
\begin{equation}
R_{\nu \rho} n^{\rho}_{A} e^{\nu}_{a} = R_{\mu \nu \sigma \rho} n^{\mu}_{B} n^{\sigma B} n^{\rho}_{A} e^{\nu}_{a} + \nabla_{a}\mathcal{K}_{A} - \nabla^{b} \mathcal{K}_{Aab} \, . \label{contracted codazzi}
\end{equation}
Finally, one may employ~\eqref{eq:completerelations} to express the normal projections of the curvature tensors as projections of the tangential basis,
\begin{equation}
R_{\mu \nu }n_{B}^{\mu }n^{\nu B} =  R -R_{\mu \nu }\gamma^{ab}e_{a}^{\mu }e_{b}^{\nu } \, ,
 \end{equation}
\begin{equation}
R_{\mu \nu \rho \sigma }n^{\mu A}n^{\nu B}n_{A}^{\rho }n_{B}^{\sigma } = R -2\gamma ^{ab}R_{\mu \nu }e_{a}^{\mu }e_{b}^{\nu } +\gamma ^{ac}\gamma ^{bd}R_{\mu \nu \rho \sigma }e_{a}^{\mu }e_{b}^{\nu }e_{c}^{\rho }e_{d}^{\sigma } \, .
\end{equation}
\section{Infinitesimal variation of submanifolds}
Consider infinitesimal variations of the surface $\Sigma$, represented as $X^{\mu} \rightarrow X^{\mu} + \delta X^{\mu}$, where the change in $X^{\mu}$ is restricted to the normal direction, i.e., $\delta X^{\mu} = \epsilon \xi^{\mu} = \epsilon \xi^{A} n_{A}^{\mu}$, with $\epsilon$ being an infinitesimal parameter. Tangential variations are ignored since they give rise to a constraint equation. This perturbation induces variations in all vectors and tensors defined on the surface $\Sigma$. In order to determine this class of variations, one has to parallel transport the displaced object back to the original point. The comparison between the two configurations, namely the parallel-transported to the original one, results into the total variation.

Applying this method to the basis vectors $e^{\mu}_{a}$, one obtains
\begin{equation}
    \delta e^{\mu}_{a} = \epsilon \left[ \left( \nabla_{a} \xi^{A} \right) n^{\mu}_{A} + \left( \xi^{A} \mathcal{K}^{b}_{a A} \right) e^{\mu}_{b} \right],
\end{equation}
where the Gauss--Weingarten equations  have been applied.

For the induced metric $\gamma_{ab}$, its variation under the infinitesimal normal deformation can be expressed in terms of the extrinsic curvatures as
\begin{equation}
    \delta \gamma_{ab} = 2 \epsilon \xi^{A} \mathcal{K}_{ab A}.
\end{equation}
From this, the variation of the determinant of the metric tensor, $\sqrt{\gamma}$, under infinitesimal normal variations of $\Sigma$, is
\begin{equation}
    \delta  \sqrt{\gamma} = \epsilon \sqrt{\gamma} \xi^{A} \mathcal{K}_{A}.
\end{equation}
The calculation of the variation of the extrinsic curvature tensors is more intricate, such that it has to be treated indirectly, following Refs. \cite{yano1978infinitesimal,chen1978theory}. For a vector field $v^{\mu}$ along the surface $\Sigma$, its variation is computed as
\begin{equation}
    \delta v^{\mu} = \bar{v}^{\mu} - v^{\mu} + \Gamma^{\mu}_{\alpha \beta} \epsilon^{\alpha} v^{\beta} \epsilon \, , \label{variationofvector}
\end{equation}
where $\bar{v}^{\mu}$ is the vector field evaluated at the deformed point $\bar{x}^{\mu}$.

Taking into account the fact that $\delta \nabla_{a} v^{\mu}$ transforms as a vector, it is straightforward to demonstrate that
\begin{equation}
    \delta \nabla_{a} v^{\mu} - \nabla_{a} \delta v^{\mu} = R^{\mu}_{\alpha \gamma \beta} e^{\beta}_{a} \xi^{\gamma} v^{\alpha} \epsilon \, .
\end{equation}
Furthermore, this result can be generalized to an arbitrary tensor $T^{\mu}_{b A}$ as
\begin{equation}
    \delta \nabla_{a} T^{\mu}_{b A} - \nabla_{a} \delta T^{\mu}_{b A}  = R^{\mu}_{\alpha \gamma \beta} e^{\beta}_{a} \xi^{\gamma} T^{\alpha}_{b A} \epsilon - \delta \Gamma^{k}_{ab} T^{\mu}_{k A} - \delta \Gamma^{B}_{a A} T^{\mu}_{b B} \, .
\end{equation}
Applying this to the basis $e^{\mu}_{b}$, we obtain the expression:
\begin{equation}
    \delta \nabla_{a} e^{\mu}_{b} - \nabla_{a} \delta e^{\mu}_{b} = R^{\mu}_{\alpha \gamma \beta} e^{\beta}_{a} \xi^{\gamma} e^{\alpha}_{b} \epsilon - \delta \Gamma^{k}_{ab} e^{\mu}_{k} \, .
    \label{deltae1}
\end{equation}
However, this expression can be calculated by using \
Eq. \eqref{eq:covariantofn}  as
\begin{align}
    \delta \nabla_{a} e^{\mu}_{b} - \nabla_{a} \delta e^{\mu}_{b} =- n^{\mu}_{A} \delta \mathcal{K}^{A}_{ab}  - \epsilon \nabla_{a} \nabla_{b} \xi^{\mu} \, , 
    \label{deltae2}
\end{align}
At this point, one may compare \eqref{deltae1} and \eqref{deltae2}, and contract with the vector $n^{A}_{\mu}$, to isolate the variation of the extrinsic curvature tensor. After rearranging terms, this variation can be expressed as
\begin{align}
    \delta \mathcal{K}^{A}_{ab} ={}& \left( - \delta^{A}_{B} \nabla_{a} \nabla_{b} + \mathcal{K}_{b B}^{c} \mathcal{K}_{ac}^{A} - R_{\mu \nu \rho \sigma} n^{\mu A} n^{\sigma}_{A} e^{\rho}_{a} e^{\nu}_{b} \right) \xi^{B} \, , \\
    \delta \mathcal{K}^{A} ={}& \left( - \delta^{B}_{A} \nabla^{a} \nabla_{a} + \mathcal{K}_{B}^{ac} \mathcal{K}_{ac}^{A} - \gamma^{ab} R_{\mu \nu \rho \sigma} n^{A \mu} n^{\sigma}_{B} e^{\rho}_{a} e^{\nu}_{b} \right) \xi^{B} \, .
\end{align}
Finally, as is noted in \cite{bhattacharyya2014entanglement}, the variation of the bulk curvature tensor and its contractions are cast in the form
\begin{align}
\delta R_{\mu \nu \rho \sigma } ={}&\epsilon n_{A}^{\alpha }\xi ^{A}{ \nabla }_{\alpha }R_{\mu \nu \rho \sigma }  \\
\delta R_{\mu \nu } ={}& \epsilon n_{A}^{\alpha }\xi ^{A}{ \nabla }_{\alpha }R_{\mu \nu } \\
\delta R ={}&\epsilon n_{A}^{\alpha }\xi ^{A}{ \nabla }_{\alpha }R
\end{align}
Putting all the variations of $L_{\Sigma}$ together leads to the extremality condition for this conformal functional
\begin{align}
& \mathcal{K}_{A} \left( W_{ab}^{ab} - P^{B}_{ab} P_{B}^{ab} \right) - \nabla^{a} \left( \nabla_{a} \mathcal{K}_{A} - 2 R_{\mu \nu \sigma \rho} n^{\mu}_{B} n^{\sigma B} n^{\rho}_{A} e^{\nu}_{a} \right) +\mathcal{K}_{B}\mathcal{K}_{ab}^{B}\mathcal{K}_{A}^{ab}  \nonumber\\
& \quad -2\mathcal{K}_{B}^{ab}\mathcal{K}_{ac}^{B}\mathcal{K}_{A b}^{c} + \gamma ^{ac}\gamma ^{bd}e_{a}^{\mu }e_{b}^{\nu }e_{c}^{\rho }e_{d}^{\sigma }n_{A}^{\alpha }{\nabla}_{\alpha}R_{\mu \nu \rho \sigma } - \gamma ^{ab}e_{a}^{\mu }e_{b}^{\nu }n_{A}^{\alpha }{ \nabla }_{\alpha }R_{\mu \nu }  \nonumber \\
& \quad +\frac{1}{3}n_{A}^{\alpha }{ \nabla }_{\alpha }R +2R_{\mu \nu \rho \sigma }n^{\mu B}n_{A}^{\rho }e_{b}^{\sigma }e_{a}^{\nu }\mathcal{K}_{B}^{ab}  -R_{\mu \nu \rho \sigma }\gamma ^{ab}n^{\mu B}n_{A}^{\rho }e_{b}^{\sigma }e_{a}^{\nu }\mathcal{K}_{B} = 0 \label{Eqs Lsigma} \, .
 \end{align}

\bibliographystyle{JHEP}
\bibliography{biblio}

\end{document}